\documentstyle {report}
\begin{document}
\pagestyle{headings}
\title{Semi-Complex Analysis  and  Mathematical Physics}
\author{Francesco Antonuccio \\
         Wadham College \\
         Oxford OX1 3PN \\
         United Kingdom \\
         email: antonucc@maths.ox.ac.uk}
\maketitle

\begin{Large}
{\centerline {\bf Author's Note}}
\end{Large}
The aim of this exposition is two-fold. First, we wish to
acquaint the reader with the semi-complex algebra\footnote{
or {\em hyperbolic quasi-numbers}; see \cite{czech}} formalism.
This is dealt with in Parts One and Two. In Part Three, we
discuss physics related topics (in particular, space-time and gravity)
in the context of this formalism.

Presently, we know that quantum mechanics finds its most natural expression
in the language of complex numbers. In particular, many fruitful
concepts arise by adopting these numbers as the basic
building blocks of the theory. In Einstein's formulation of
 General Relativity,
the essential building blocks are the real numbers, which are used in the
construction of real manifolds (with curvature).

It seems, then, that the  starting point of each theory differs
in a fundamental way, since one is deeply rooted in the theory of
complex numbers and unitary transformations, while the other makes no explicit
mention of these concepts.
One approach towards reconciling
these two theories
involves re-expressing the relevant parameters in general relativity in terms
of complex valued quantities (see, for example, \cite{Roger}).

Rather than making use of complex numbers directly,
the approach adopted in this exposition involves constructing
a formalism  which {\em looks like} the usual complex algebra formalism.
More precisely, we perform a commutative extension of the real numbers,
but instead of arriving at the division algebra of complex numbers, we
arrive at a  {\em non}-division algebra, which naturally admits `conjugation'
of elements.

Adopting these numbers as our building blocks produces a theoretical
framework which is intrinsically non-euclidean in character.
Connections can then be made between the symmetries of space-time,
and the complex unitary groups that arise in particle gauge theories.

\medskip
{\bf Acknowledgements}

This work was entirely supported by the Commonwealth
Scholarship and Fellowship Plan (The British Council, U.K.).

\newcommand{\cn}{\overline}
\newcommand{\D}{${\bf D}$}
\newcommand{\beq}{\begin{equation}}
\newcommand{\eeq}{\end{equation}}
\newcommand{\ifff}{\Leftrightarrow}
\newcommand{\defn}{ \stackrel{\rm def}{=}}
\newcommand{\dw}{\Delta w}
\newcommand{\db}{\partial}
\newcommand{\Dn}{{\bf D}^n}
\newcommand{\Dm}{{\bf D}^m}
\newpage
\tableofcontents
\newtheorem{proposition}{Proposition}[chapter]
\newtheorem{theorem}{Theorem}[chapter]
\newtheorem{definition}{Definition}[chapter]
\newtheorem{corollary}{Corollary}[chapter]
\newtheorem{example}{Example}[chapter]
\newpage
\part{Semi-Complex Analysis in One Variable}    \label{onevariable}
\newpage


%
%



%
%




\chapter{The Semi-Complex Number System}
\section{Definition and Terminology}
An informal yet instructive way of introducing the complex number system
{\bf C} to a newcomer is to postulate the existence of `numbers' having the
form
\begin{equation}
 z = x+iy
\end{equation}
where $x$ and $y$ are  real numbers and $i$ is a
commuting  variable
satisfying the relation
\begin{equation}
  i^2 = -1.  \label{isq}
\end{equation}
If we write $i^2 = +1$ instead of $i^2 = -1$, what kind of number system
do we end up with? To explore this possibility we will consider
`numbers' of the form
\begin{equation}
\mbox{ \fbox{$w=t+jx$}}    \label{scn}
\end{equation}
where $t$ and $x$ are real and $j$ is a commuting
 variable satisfying the relation
\begin{equation}
\mbox{\fbox{$j^2=+1$}}      \label{j}
\end{equation}
The algebra defined by (\ref{scn}) and
(\ref{j}) will be denoted by the symbol {\bf D}, and
we will often reserve the letter $w$
  for  elements in {\bf D}.
This number system turns out to be a peculiar creature, which I have ventured
to call the {\it semi-complex number system}. The precise reason
for this will become clear as this exposition unfolds.

We can add and multiply any two semi-complex numbers by assuming
(tentatively) that the usual rules of arithmetic apply, as well as
stipulating that $j^2 = +1$. Doing so enables us to write down
immediately the rules for addition and multiplication:

\noindent
{\it Addition}:
\begin{equation}
(t_1+jx_1)+(t_2+jx_2) = (t_1+t_2)+j(x_1+x_2)
\end{equation}
\noindent
{\it Multiplication}:
\begin{equation}
(t_1+jx_1)\cdot (t_2+jx_2) = (t_{1}t_{2}+x_{1}x_{2})+j(t_{1}x_{2}+x_{1}t_{2})
\end{equation}
We are now in a position to give a formal definition of the
semi-complex algebra \D; namely, {\D}  is the set of {\em ordered
pairs} of real numbers $(t,x)$, equipped with the following operations of
addition and multiplication:
\begin{enumerate}
 \item $(t_1,x_1)+(t_2,x_2) = (t_1+t_2,x_1+x_2)$
 \item
 $(t_1,x_1)\cdot (t_2,x_2) = (t_{1}t_{2}+x_{1}x_{2},t_{1}x_{2}+x_{1}t_{2})$
\end{enumerate}
The reader may like to check that the above rules do indeed yield an
associative, distributive and commutative algebra.

With these definitions, it is easy to see that an element of the form
$(t,0)$ can be identified with the real number $t$, and $(0,1)$
can be identified with $j$, since
\[ (0,1) \cdot (0,1) = (1,0). \]
Also,
\[ (0,1) \cdot (x,0) = (0,x), \]
so we may legitimately identify the ordered pair $(t,x)$ with our
old friend $t+jx$.
At any rate, we  unabashedly write
\[ w = t+jx \]
for any semi-complex number $w\in${\bf D}, and it is this notation
that we will adopt throughout. Incidentally,
we choose to call  $t$ and $x$  respectively
  the {\it real} and {\it imaginary} parts
of  $w= t+jx$, where the overlap of terminology with the complex
numbers is intentional.

\section{Conjugation and Inversion}
Of course, now that we have taken the liberty to view {\D} as
a `number system', it remains to show that we can do more
than just add and multiply different elements. In particular,
we would like to know which elements in {\D} have inverses.

First, we introduce the notion of {\em conjugation}; given
any semi-complex number $w=t+jx$, the {\em conjugate} of $w$,
written $\cn{w}$, is defined to be
\[ \cn{w} = t-jx . \]
Two simple consequences can be immediatetely deduced; for
any $w_1,w_2 \in$ \D, we have
\begin{eqnarray}
 \overline{w_1+w_2} & = & \overline{w_1}+\overline{w_2} \hspace{4mm}
                           \mbox{and} \\
 \overline{w_1 \cdot w_2} & = & \overline{w_1} \cdot \overline{w_2}.
\end{eqnarray}
We also have the important identity
\beq
\cn{w} \cdot w = t^2 - x^2.
\eeq
Hence $\cn{w} \cdot w$ is {\em real} for any semi-complex
number $w$, although unlike the complex case, it may
take on {\em negative } values. In order to strengthen the analogy
between the semi-complex and complex numbers, we often write
\[ |w|^2 = \cn{w} \cdot w  \]
where $|w|^2$ is referred to as the `modulus squared'
of $w$. A nice  consequence of these definitions
can now be stated:
\begin{proposition}
For any semi-complex numbers $w_1$, $w_2 \in$ {\bf D},
\[ |w_1 \cdot w_2|^2 = |w_1|^2 \cdot |w_2|^2 . \] \label{propmult}
\end{proposition}
{\bf Proof}: Exercise! \hfill $\Box$

Now observe that if $|w|^2$ does not vanish, the quantity
\beq
  w^{-1} = \frac{\overline{w}}{|w|^{2}}
\eeq
is a well defined inverse for $w$. So $w$ fails to have an
inverse if (and only if)
\beq
 |w|^2 = t^2-x^2 = 0, \label{mod0}
\eeq
or simply when $x= \pm t$. If we view $t$ as time and let $x$ be a space
coordinate, relation (\ref{mod0}) defines the light cone in a $1+1$
spacetime ($c$ = velocity of light = 1). Using this jargon, the
number $w=t+jx$ fails to have an inverse if the spacetime point $(t,x)$
lies on the light cone.

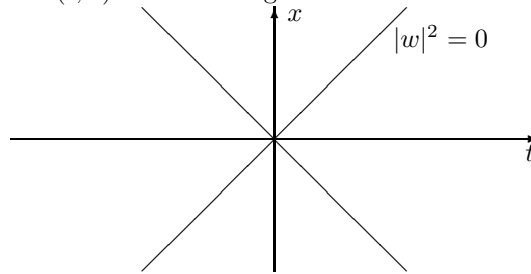
\begin{picture}(280,100)(0,0)
 \put(40,50){\vector(2,0){200}}
 \put(140,0){\vector(0,1){100}}
 \put(90,0){\line(1,1){100}}
 \put(90,100){\line(1,-1){100}}
 \put(235,42){$t$}
 \put(145,95){$x$}
 \put(185,85){$|w|^2 = 0$}
\end{picture}

\medskip
An important distinction between the complex and semi-complex numbers can
now be discerned: {\bf D} fails to be a division algebra. On first impression,
this seems to cast a dark shadow upon the respectability of {\bf D}
as a potentially useful number system. However, we
have just seen that those elements in {\D} that are not invertible
can be assigned  special physical significance. This suggests that focusing
exclusively on
number systems that are division algebras may be unnecessarily
restrictive, especially if we are seeking to construct physical
theories out of them.

\section{The Order Properties of {\bf D} (Optional)}
\label{orderproperties}
In this section (which may be omitted on a first reading)
 we discover that the semi-complex algebra
possesses certain order properties analogous to those of the
real numbers, and, moreover, that these  are related
to the causal structure of {\D} viewed as $1+1$ space-time.

\medskip
For any semi-complex number $w=t+jx$, we introduce two associated
{\em real} quantities, $w_{+}$ and $w_{-}$, by writing
\beq
w_{+} = t+x \hspace{4mm} \mbox{and} \hspace{4mm} w_{-} = t-x.
\eeq
A straightforward calculation yields the identities
\beq
 (w_{1}\cdot w_{2})_{\pm} = (w_{1})_{\pm} \cdot (w_{2})_{\pm} \label{pmproduct}
\eeq
The order relation $\leq$ on {\D} is defined by writing
\[
      w_1 \leq w_2 \hspace{4mm} \mbox{(or equivalently, $w_2 \geq w_1$)} \]
whenever
\[ (w_{1})_{+} \leq (w_{2})_{+} \hspace{4mm} \mbox{and} \hspace{4mm}
    (w_{1})_{-} \leq (w_{2})_{-} . \]
Of course, if we replace every occurrence of the symbol $\leq$ in the
above definition
 with the strict inequality $<$ (or $>$), we have the corresponding
definition for the meaning of the relation $w_1 < w_2$ (or $w_2 > w_1$).
Notice that these definitions give the usual meanings for $\leq$ and $<$
when restricted to the real numbers.

Three distinctive properties of  $\leq$ can be immediately deduced:
\begin{enumerate}
 \item {\em Reflexivity}: $ w \leq w$ for any $w\in$ \D;
 \item {\em Anti-Symmetry}:
                       If $w_1 \leq w_2$ and $w_2 \leq w_1$, then $w_1 = w_2$;
 \item {\em Transitivity}:
 If $w_1 \leq w_2$ and $w_2 \leq w_3$, then $w_1 \leq  w_3$;
\end{enumerate}
These properties imply that the semi-complex algebra (equipped with
the relation $\leq$) forms a partially ordered set (or {\em poset}).
However, we have additional properties, which we  list in the next
proposition:
\begin{proposition}
 Let $a,b,c,d$ be elements in \D.
\begin{enumerate}
 \item If $a \leq b$, then $a+c \leq b+c$;
 \item If $a \leq b$ and $c \leq d$, then $a+c \leq b+d$;
 \item If $a \leq b$ and $c \geq 0$, then $a\cdot c \leq b \cdot c$;
 \item If $a \leq b$ and $c \leq 0$, then $b \cdot c \leq a \cdot c$;
 \item If $a > 0$, then $1/a > 0$;
 \item If $a < 0$, then $1/a < 0$.
\end{enumerate}
\end{proposition}
{\bf Proof}: Use the identities (\ref{pmproduct}). \hfill $\Box$

\medskip
For the real numbers, we may define the subset of
strictly positive numbers $P = \{ a \in {\bf R}| a>0 \}$.
Likewise, we may define the set $C_{+}$ of {\em strictly positive}
semi-complex numbers by setting
\beq
    C_{+} = \{ w \in {\bf D} | w>0 \}.
\eeq
As is the case for the subset $P$ of strictly positive reals,
 the subset $C_{+}$ is closed
under the operations of addition, multiplication, and inversion.

The subset $C_{+}$ has a nice physical interpretation; one can see from
the definitions that it consists of those points in space-time
lying inside the  future light cone emanating from the origin.

Note that if $|w|^2 \neq 0$, then precisely {\em one} of the following
numbers,
        \[ w, -w, jw, -jw, \]
                      lies in $C_{+}$. We may now define
a function which sends any $w\in${\D} to a corresponding
element in the set of `non-negative' numbers $\{w| w \geq 0 \}$ by setting
\beq
  |w|_{\ast} = \left\{ \begin{array}{cl}
                         w & \mbox{if $w\geq 0$} \\
                        -w &  \mbox{if $-w\geq 0$} \\
                        jw &  \mbox{if $jw\geq 0$} \\
                        -jw &  \mbox{if $-jw\geq 0$}
                        \end{array}
                \right.
\eeq
The reader may like to check that this is a well defined
function on {\D}, and moreover, gives rise to the following
identities
\begin{eqnarray}
         |w_1 + w_2|_{\ast} & \leq & |w_1|_{\ast} + |w_2|_{\ast} \\
          |w_1 \cdot w_2|_{\ast} &  = & |w_1|_{\ast} \cdot |w_2|_{\ast}
\end{eqnarray}
Of course, restricting $|\cdot|_{\ast}$ to the reals yields the
usual `absolute value' function on ${\bf R}$.


\section{The Semi-Complex Norm (Optional)}
In this section (which can be omitted on a first reading) we
show that one can define a norm on the semi-complex
algebra which is analogous to the complex modulus.

To begin, consider  the two functions
$| \cdot|_{\pm} : {\bf D}\rightarrow [0.\infty)$ defined by
\begin{eqnarray}
 |t+jx|_{+} & = & |t+x| \\
 |t+jx|_{-} & = & |t-x|
\end{eqnarray}
Note that $w=t+jx$ is invertible if and only if both $|w|_{+}$
and $|w|_{-}$ are non-zero. It is now a straightforward exercise
to check that, for any $w,w_1,w_2 \in {\bf D}$,
\begin{enumerate}
 \item $|w|_{\pm} \geq 0 $
 \item $|w_1 + w_2|_{\pm} \leq |w_1|_{\pm} + |w_2|_{\pm}$
 \item $|w_1 \cdot w_2|_{\pm} = |w_1|_{\pm} \cdot |w_2|_{\pm}$
 \item $|w|_{\pm} = 0 \ifff w = 0$
\end{enumerate}
where the notation $|w|_{\pm} = 0$ means $|w|_{+} = 0$ {\em and}
$|w|_{-} = 0$.
So  the two functions $|\cdot|_{+}$ and $|\cdot|_{-}$ are semi-norms
on \D.
Having defined these semi-norms , we are now  in a position
to define the semi-complex norm
$|| \cdot|| : {\bf D}\rightarrow [0.\infty)$; namely, for any
 $w=t+jx$, we write
\begin{eqnarray}
 ||w|| & = & \sqrt{ |w|_{+}^2 + |w|_{-}^{2}} \label{seminorm} \\
       & = & \sqrt{2(t^2+x^2)} \label{norm2}
\end{eqnarray}
 The properties of   $|\cdot|_{+}$ and $|\cdot|_{-}$ listed earlier
can be used to prove the following;
for any $w,w_1,w_2\in {\bf D}$,
\begin{enumerate}
 \item $||w|| \geq |w|_{\pm} \geq 0$;
 \item $||w^n|| = 0 \ifff w=0$ for any positive integer $n$;
 \item $||w_1 + w_2|| \leq ||w_1|| + ||w_2||$;
 \item $||w_1 \cdot w_2|| \leq ||w_1|| \cdot ||w_2||$.
\end{enumerate}
{\bf Proof}: Easy exercise! \hfill $\Box$

\medskip

In order to define limits in {\D} , we proceed
as in the case for complex numbers, except we replace the complex
modulus (or norm) by the semi-complex norm defined above. The induced
topology is just the familiar topology for ${\bf R}^2$.


\chapter{The Calculus of Semi-Complex Functions}
\section{Some Elementary Functions}
We define the {\em exponential function} $e^w$ by invoking
a familiar series expansion;
\beq
       e^{w} \defn 1+w+\frac{w^2}{2!} + \frac{w^3}{3!}+ \cdots , \label{exp}
\eeq
which converges  for all $w$ in the semi-complex (i.e. $t$-$x$ ) plane.
Since the elements in {\D} commute, we have, for any $w_1,w_2 \in {\bf D}$,
the identity
\beq
           e^{w_1} \cdot e^{w_2} = e^{w_1 +w_2}. \label{e1e2}
\eeq
For real $\theta$, we have
\beq
e^{j\theta} = \cosh{\theta} + j\sinh{\theta},
\eeq
which can be derived by substituting $w=j\theta$ into
the expansion (\ref{exp}). So, in terms of the real
and imaginary parts of $w$, the exponential $e^w$
takes the form
\beq
e^{w} = e^{t}(\cosh{x} + j\sinh{x}).
\eeq
Of course, this last identity may be taken as our
defining expression for the exponential function.

Notice that for real values of $\theta$,
\[ |e^{j\theta}|^2 = \cosh^2\theta - \sinh^2\theta = 1.\]
In the light of proposition \ref{propmult}, we may view
$e^{j\theta}$ as a `phase factor'
which leaves the modulus squared of a semi-complex number unchanged
after multiplication. This is of course analogous to the complex phase
factor $e^{i\theta}$ encountered in complex number theory.

\medskip
Our next task is to define the {\em logarithmic function} $\log w$.
Let $w = t+jx  > 0$ be a given {\em strictly positive} semi-complex number
(see Section \ref{orderproperties} for a discussion on the
use of inequalities in \D ). Then the expression
\beq
\log{w} = \frac{1}{2}\ln{(t^2-x^2)} + j\tanh^{-1}{(\frac{x}{t})},
                          \label{logarithm}
\eeq
ensures that the identity
\beq
 e^{\log{w}} = w
\eeq
holds, and, moreover, it is the {\em unique} such expression.
Using this fact, and identity (\ref{e1e2}), we arrive at a familiar looking
result:
\beq
 \log w_1 w_2 = \log w_1 + \log w_2  \hspace{5mm}
                                      \mbox{ for any $w_1,w_2 > 0$.}
\eeq
The reader is reminded that the condition $w = t+jx >0$ implies
that the space-time point $(t,x)$ lies inside the {\em future} light cone
with vertex at the origin.

\section{Differentiation and Holomorphic Functions}
\subsection{The Derivative}
For an arbitrary {\bf D}-valued function $f$ defined on some open
region $U$ of the semi-complex plane, we may write
\[ f(t+jx) = u(t,x)+jv(t,x) , \hspace{6mm} t+jx \in U , \]
where $u$ and $v$ are real valued functions.
 Unless otherwise stated, we will always assume
$u$ and $v$ to be {\em smooth} (or $C^{\infty}$) functions.

\medskip
Defining the derivative of a function at some point usually
involves investigating the behaviour of the quotient
\beq
    \frac{f(w+\dw) - f(w)}{\dw}  \label{quotient}
\eeq
as $\dw$ `tends to zero'. An alternative formulation of the
derivative which will be convenient for our purposes is based
on the `linear approximation'
\[ \Delta f = f(w+\dw) - f(w) \approx f'(w)\cdot \dw, \]
which becomes `exact' as $\dw$ becomes vanishingly small.
Adopting this last point of view enables us to define the
derivative of a semi-complex function without introducing the
need to  divide by the semi-complex parameter $\dw$, which
arises in the quotient (\ref{quotient}).

\begin{definition} Let $f: U \rightarrow {\bf D}$ be a semi-complex
 function defined on some open region $U$ of the semi-complex plane.
 Let $w_0 \in U$. Then $f'(w_0)\in {\bf D}$ is said to be the
 {\em derivative} of f at $w_0$ if the quotient
\[
 \frac{ f(w_0+\dw) - f(w_0) - f'(w_0)\cdot \dw }{||\dw||}
\]
tends to zero as $||\dw|| \rightarrow 0$.

\medskip
If $f$ has a derivative  at each point
in its domain $U$, it is said to be {\em holomorphic} on $U$.
\end{definition}

\medskip
As in the theory of complex functions, the real and imaginary parts
of a holomorphic semi-complex  function $ u+jv$  satisfy certain
relations. In the complex case, these are known as the
Cauchy-Riemann Equations.  To discover what the corresponding
 relations are in the semi-complex case, we
observe that the existence of a derivative for  $f$ at $w\in U$
implies
\beq
f'(w) = \lim_{h \rightarrow 0} \frac{f(w+h)-f(w)}{h} =
           \lim_{h \rightarrow 0} \frac{f(w+jh)-f(w)}{jh},
\eeq
where $h$ is a {\em real} parameter.
Hence, using the notation $f = u+jv$, we deduce that
\beq
f' = \frac{\partial{u}}{\partial{t}}+j\frac{\partial{v}}{\partial{t}}
= \frac{\partial{v}}{\partial{x}}+j\frac{\partial{u}}{\partial{x}}.
\eeq
Equating real and imaginary parts above, we arrive at the
necessary condition for holomorphicity:
\beq
\frac{\partial{u}}{\partial{t}}=\frac{\partial{v}}{\partial{x}}, \hspace{1cm}
\frac{\partial{u}}{\partial{x}}=\frac{\partial{v}}{\partial{t}}, \label{CR}
\eeq
which are analogs of the Cauchy-Riemann equations arising in complex
analysis.
An easy calculation shows that these relations  force $u$ and $v$
(and thus $f$) to satisfy the one dimensional wave equation:
\[ \frac{\partial^{2}{u}}{\partial{t^2}}- \frac{\partial^{2}{u}}{\partial{x^2}}
= 0, \hspace{1cm}\frac{\partial^{2}{v}}{\partial{t^2}}
-\frac{\partial^{2}{v}}{\partial{x^2}}=0. \]
These observations suggest that holomorphic semi-complex functions
might play a useful role in special relativity -- a speculation that we will
pursue in greater detail later on.

\medskip
We now wish to prove that a semi-complex function with real and
imaginary parts satisfying  relations (\ref{CR}) in some open
region $U$  is holomorphic.
The argument follows very closely the one given in \cite{ahlfors} for
the analogous situation in complex function theory.
{}From the calculus, we can write
\begin{eqnarray*}
 u(t+ \Delta t, x + \Delta x) - u(t,x) & = & \frac{\db u}{\db t} \Delta t
    + \frac{\db u}{\db x} \Delta x + \epsilon_1 , \\
 v(t+ \Delta t, x + \Delta x) - v(t,x) & = & \frac{\db v}{\db t} \Delta t
    + \frac{\db v}{\db x} \Delta x + \epsilon_2 ,
\end{eqnarray*}
where the remainders $\epsilon_1, \epsilon_2$ tend to zero more
rapidly than $\dw = \Delta t + j\Delta x$, in the sense that
$ \epsilon_1 / || \dw||$ and $ \epsilon_2 / || \dw|| $ tend to zero
as $||\dw||  \rightarrow 0$.

With the notation $f(w) = u(t,x) +jv(t,x)$, we obtain by virtue of the
relations (\ref{CR})
\beq
 f(w+\dw)-f(w) =
\left( \frac{\partial{u}}{\partial{t}}+j\frac{\partial{v}}{\partial{t}}
    \right) \cdot \dw + \epsilon_1 + j\epsilon_2,
\eeq
and therefore the quotient
\[
\frac{ f(w+\dw)-f(w) -
\left( \frac{\partial{u}}{\partial{t}}+j\frac{\partial{v}}{\partial{t}}
    \right) \cdot \dw }{||\dw ||} = \frac{\epsilon_1 + j\epsilon_2}{||\dw||}
\]
tends to zero as  $||\dw||  \rightarrow 0$.
We conclude that $f(w)$ is holomorphic (on $U$).

\medskip
\subsection{Semi-Complex Holomorphic Functions}
Having shown that relations (\ref{CR}) are the {\em necessary
and sufficient} conditions for holomorphicity,
 we may approach the whole subject
of semi-complex holomorphic functions without any explicit reference
to limiting procedures at all.

Let us first introduce the operators
\begin{equation}
\frac{\partial}{\partial{w}} = \frac{1}{2}\{\frac{\partial}{\partial{t}}+
j\frac{\partial}{\partial{x}}\} ,\hspace{1cm}
 \frac{\partial}{\partial{\overline{w}}} = \frac{1}{2}\{
\frac{\partial}{\partial{t}}-j\frac{\partial}{\partial{x}}\}.
\end{equation}
A straightforward calculation shows that
\begin{equation}
df = \frac{\partial{f}}{\partial{w}}dw + \frac{\partial{f}}{\partial{
\overline{w}}}d\overline{w}, \label{df}
\end{equation}
for any semi-complex valued function $f$. This result is particularly
useful, since it is tantamount to saying that the variables $w$
and $\overline{w}$ may be treated  {\em independently}, a fact which
will be well appreciated in later sections of this article.

\medskip
Now consider the constraint equation
\begin{equation}
\frac{\partial{f}}{\partial{\overline{w}}} = 0. \label{dhol}
\end{equation}
Separating real and imaginary parts, condition (\ref{dhol}) takes the form
\begin{equation}
\frac{\partial{u}}{\partial{t}}=\frac{\partial{v}}{\partial{x}}, \hspace{1cm}
\frac{\partial{u}}{\partial{x}}=\frac{\partial{v}}{\partial{t}}, \label{dcr}
\end{equation}
which are precisely the relations (\ref{CR}) satisfied
by a semi-complex holomorphic function. This enables us to {\em define}
holomorphic functions as precisely those for which the constraint equation
(\ref{dhol}) holds. (Compare this with the definition of a {\em complex }
holomorphic function).

\medskip
By virtue of the holomorphicity condition (\ref{dhol}), the identity
(\ref{df}) simplifies to
\begin{equation}
 df = \frac{\partial{f}}{\partial{w}}dw,  \label{dfhol}
\end{equation}
which shows that the derivative $f'$ of a {\em holomorphic} function $f$
(sometimes written $df/dw$) may be defined as
\begin{equation}
f' = \frac{\partial{f}}{\partial{w}} .
\end{equation}
A simple calculation demonstrates  that $f'$ above has the explicit form
\begin{equation}
f' = \frac{\partial{u}}{\partial{t}}+j\frac{\partial{v}}{\partial{t}}
= \frac{\partial{v}}{\partial{x}}+j\frac{\partial{u}}{\partial{x}},
\end{equation}
which is of course  consistent with our ealier observations.
\subsection{Derivatives of Elementary Functions}
Fortunately, the results just obtained  yield familiar looking results.
For example,
the reader is invited to verify the following fundamental facts;
\begin{equation}
\frac{d}{dw}w^{n}=nw^{n-1}, \hspace{.8cm}
\frac{d}{dw}e^{w}=e^{w},\hspace{.8cm}
\frac{d}{dw} \log{|w|_{\ast}} = \frac{1}{w},
\end{equation}
where the first two functions above are defined everywhere,
and the logarithm is defined for all $w$ satisfying $|w|^2 \neq 0$.

\medskip
We remark that the holomorphic map $f:(t,x) \longrightarrow (u,v)$ induces
a conformal transformation on the semi-complex plane in the following sense:
\begin{equation}
du^2-dv^2 = |f'|^{2}(dt^2-dx^2),
\end{equation}
which is a direct consequence of (\ref{dfhol})
(take the modulus of both sides).
The conformal factor is thus $|f'|^2$. The coordinates $(u,v)$ define
a coordinate system which possesses the natural property that the
coordinate curves ($u=$ const, $v=$const) intersect orthogonally with
respect to the Lorentz inner product with signature $(+,-)$. This is of course
perfectly analogous to the `isothermal coordinates' one encounters in
complex analysis.

\medskip
Integration of semi-complex valued functions offers additional insights; we
pursue this topic next.

\section{Semi-Complex Integeration}
\subsection{Definition}
Integrating a semi-complex function $f = u+jv$  along a piecewise smooth
curve $\gamma$
is defined in the obvious way; namely, we set
\begin{eqnarray*}
 \int_{\gamma} f dw & = & \int_{\gamma} (u+jv)(dt +jdx) \\
                    & = & \int_{\gamma} udt + vdx +
                           j \int_{\gamma} vdt + udx ,
\end{eqnarray*}
where the last line above involves the evaluation of two line integrals
in the semi-complex  plane. We will always assume that the path of integration
is  piecewise smooth.

\subsection{Cauchy Type Theorems}
If the path of integeration is a simple closed curve, then
under certain favourable conditions, we may invoke Green's
Theorem in the plane to convert the line integral into
a surface integral. This is the content of the next Proposition;
\begin{proposition}
Let $f$ be a semi-complex function defined on and within
a smooth, simple-closed curve $C$. Let $R$ denote the interior of $C$. Then
\[ \oint_{C} f dw = \int\!\!\int_{R}
\frac{\partial{f}}{\partial{\overline{w}}} d\overline{w}\wedge dw . \]
\label{prop: green}
\end{proposition}
{\bf Proof}: Write $f=u+jv$, $dw=dt+jdx$, and use Green's Theorem
in the plane.\hfill  $\Box$

Of course, the last Proposition still holds for regions $R$
that are not simply connected. The details are left to the reader.
An immediate corollary of Proposition \ref{prop: green} turns out to be
formally identical to Cauchy's Theorem in complex analysis;
\begin{corollary}
Let $f$ be a semi-complex  function which is holomorphic
on and within a simple closed curve $C$.
Then
 \beq
      \oint_{C} f(w) dw = 0.
 \eeq
\end{corollary}
{\bf Proof}: Recall that $f$ is holomorphic $\ifff \frac{\db f}{\db \cn{w}}=0$.
             Then use Proposition \ref{prop: green}. \hfill $\Box$

\medskip
So intgerals of {\em holomorphic} functions along paths lying in the domain
of holomorphicity depend only on the endpoints of those paths;
consequently, the integrals are {\em path independent}. In fact,
if $F(w)$ is the anti-derivative\footnote{i.e. $F'(w) = f(w)$; all holomorphic
functions possess an anti-derivative which is unique up to a constant.}
of a holomorphic function $f(w)$, we may evaluate the integral
\[ \int_{\gamma} f(w) dw \]
by calculating the values for $F$ on the endpoints of the path $\gamma$.

For example, suppose $\gamma$ is any path from $w_1$ to $w_2$. Then
\begin{eqnarray*}
 \int_{\gamma} w^2 dw & = & \left[ \frac{w^3}{3} \right]_{w_1}^{w_2} \\
                      & = & \frac{w_2^3}{3} - \frac{w_1^3}{3}.
\end{eqnarray*}

\medskip
We now seek a result which is analogous to Cauchy's Integral Formula.
This last classical result implies that if the values of a complex
holomorphic function are known on a circle,  the values of the
function inside the circle are {\em uniquely} determined.
Now a circle in the complex plane (centered at the origin, say)
is the locus of points $z \in {\bf C}$ satisfying $|z|^2 = a^2$
(where $a$ is a real constant). For the semi-complex case,
the analogous object is the locus of points $w \in${\D} satisfying
$|w|^2 = a^2$, which is a {\em hyperbola} in the semi-complex plane.

Thus we anticipate that a semi-complex holomorphic function is
{\em uniquely} determined  everywhere\footnote{almost; if $a = 0$, then
$f$ is uniquely defined for {\em all} points in {\D}; otherwise it
is uniquely defined for all points $|w|^2 \neq 0$.}
by its values  on the hyperbola
$|w|^2 = t^2 - x^2 = a^2$. This is indeed the case, which
can be deduced from the following Proposition:
\begin{proposition}
Suppose $f$ is semi-complex holomorphic on an open set
$U \subseteq {\bf D}$. Let $\alpha = (1+j)/2$. Then
for any $p,q \in {\bf R}, w\in U$ for which
$w+p\alpha$ and $w+q\cn{\alpha}$ lie in $U$, we have
\beq
 f(w) = \cn{\alpha}f(w+p\alpha) + \alpha f(w+q\cn{\alpha}).
\eeq                        \label{CauchyIntegral}
\end{proposition}
{\bf Proof}: Let $w\in U$ be fixed.
Set $F(p,q) = \cn{\alpha}f(w+p\alpha) + \alpha f(w+q\cn{\alpha})$.
Using the fact $\cn{\alpha} \alpha = 0$, and the holomorphicity of $f$,
we deduce
\[ \frac{\db F}{\db p} = \frac{\db F}{\db q} = 0. \]
So $F(p,q)$ is constant. But $F(0,0) = f(w)$, and so the proof is complete.
\hfill $\Box$

\medskip
As a consequence, we observe
that if $f$ is holomorphic in the entire semi-complex
plane, then it is uniquely determined by its values on the light cone
$|w|^2 = 0$. If $f$ is is known to be holomorphic everywhere {\em except}
on the light cone, then it is uniquely determined by its values
on the hyperbola $|w|^2 = a^2, a\neq 0$. The verification of these facts
is a simple  `proof by picture' argument. The reader may wish to
fill in the details!

\subsection{A Counter Example}
 The reader may recall that Cauchy's Integral
Formula plays a crucial role in the proof of the analyticity property of
complex holomorphic functions. In this section we show by way of a
counter example that
{\em semi-complex holomorphic functions are not necessarily
analytic}.
Thus the  holomorphicity condition imposed on  semi-complex valued
functions turns out to be
a  much less restrictive condition than for the complex case.

\medskip
Fortunately, we don't have to look too hard to discover
an example\footnote{noticed by Steve Lack (Cambridge) and the author.}
of a non-analytic function which is holomorphic in the
entire ${\bf D}$
plane. First, we introduce the function
\begin{equation}
f(w) = e^{-\frac{1}{w^2}}, \label{nah}
\end{equation}
which is defined and holomorphic for all $w$ satisfying $|w|^2 \neq 0$. What
happens
as $|w|^2$ approaches zero (i.e. as $w$ approaches the null lines
$t= \pm x$) ? A straightforward calculation shows that
\begin{equation}
e^{-\frac{1}{w^2}} = \frac{1}{2}(1+j)e^{-\frac{1}{(t+x)^2}} +
                 \frac{1}{2}(1-j)e^{-\frac{1}{(t-x)^2}}. \label{emws}
\end{equation}
Evidently, as $t \rightarrow x$, the second term in (\ref{emws}) vanishes,
while the first is finite. Similarly, the function $e^{-1/w^2}$ is well behaved
as $t \rightarrow -x$. The whole expression vanishes as $w$ approaches zero.
These observations permit us to extend the function $e^{-1/w^2}$ to the
entire ${\bf D}$ plane in the following way: Let
\[
g(w) = \left\{\begin{array}{ll}
              e^{-\frac{1}{w^2}} & \mbox{if $|w|^2 \neq 0$} \\
              \frac{1}{2}(1 \pm j) e^{-\frac{1}{4t^2}} & \mbox{ if $t=\pm x$,
                                                                 $t\neq 0$} \\
               0 & \mbox{if $w=0$}
              \end{array}
       \right.
\]
It is now a straightforward (though somewhat tedious) exercise to show
that $g:{\bf D} \longrightarrow {\bf D}$ is holomorphic everywhere.
Note that $g$ is non-invertible on the null lines.

\chapter{The Physics of Semi-Complex Numbers}
\label{physics}
\section{The Lorentz Transformation}
It turns out that  Lorentz's Transformation on $1 + 1$
space-time
 involving a velocity
boost can be elegantly  expressed in terms of semi-complex
numbers.
The Lorentz Transformation\footnote{where the appearance of
the constant $c$ is now
made explicit}
 between two coordinate systems $(t',x')$ and
$(t,x)$ is given by the equations
\begin{equation}
ct' = \frac{ ct -\frac{v}{c}x}{\sqrt{1-\frac{v^2}{c^{2}}}},
\hspace{.85cm} x' = \frac{x-\frac{v}{c}ct}{\sqrt{1-\frac{v^2}{c^2}}}.
\label{Lorentz}
\end{equation}
If we define $\theta$ by setting
\begin{equation}
\tanh{\theta}= \frac{v}{c}, \label{phvel}
\end{equation}
then the identity $\cosh^2{\theta}-\sinh^2{\theta} =1 $ enables us to write
\begin{equation}
\cosh{\theta} = \frac{1}{\sqrt{1-\frac{v^2}{c^2}}},\hspace{.85cm}
\sinh{\theta} = \frac{v/c}{\sqrt{1-\frac{v^2}{c^2}}}.
\end{equation}
Consequently, (\ref{Lorentz}) is equivalent to the expression
\begin{equation}
ct' + jx' = (\cosh{\theta} - j\sinh{\theta})(ct +jx),
\end{equation}
or simply
\begin{equation}
\mbox{\fbox{$w' = e^{-j\theta}\cdot w$}}  \label{lore}
\end{equation}
where $w$ and $w'$ are semi-complex numbers whose real and imaginary parts
are the the time and space variables (respectively) of
the corresponding coordinate system.
So if a space-time point $(ct,x)$ is represented by the semi-complex number
$w= ct + jx$, the Lorentz Transformation in (\ref{Lorentz})  simply involves
multiplying $w$ by a {\em phase factor} $e^{-j\theta}$. A physical
theory which is required to be Lorentz invariant must therefore be
invariant under the global phase transformation
\begin{equation}
\mbox{\fbox{ $w \rightarrow w' = e^{-j\theta} \cdot w$} \hspace{7mm}
($\theta =$  constant)}
\label{global}
\end{equation}
The simplest physical theory which is invariant under this transformation
will be studied next.

\medskip
We should remark that our reformulation of Lorentz's Transformation in terms
of semi-complex numbers involves only {\em one} spatial coordinate,
which obviously falls short of the three spatial dimensions that are
known to exist. In order to handle transformations involving
multiple space dimensions, we need to introduce higher dimensional
structures that will appear formally identical to the concept of
 vector spaces and curved manifolds in ordinary analysis. We pursue
these topics in Part \ref{severalvariables}.

\section{Classical Point Particles}
Let us consider a classical point particle with rest mass
$m$ moving in $1+1$ space-time. A suitable Lagrangian
 (see \cite{goldstein})  for describing such a particle
is given by
\beq
 L = m \left[ \left( \frac{dct}{d\tau}
\right)^2
                               - \left(\frac{dx}{d\tau}\right)^2 \right],
\eeq
or simply
\begin{equation}
 L = m \frac{d\overline{w}}{d\tau} \cdot \frac{dw}{d\tau}, \label{lag2}
\end{equation}
where  $w= ct+jx$, and $\tau$ is the proper time\footnote{In fact
 $d\tau = \sqrt{d\cn{w} dw} = \sqrt{c^2dt^2-dx^2}$, and has the dimensions of
length.} , which is an invariant parameter of the trajectory.

 Note that the Lagrangian
is always real valued, and invariant under the global phase
transformation (\ref{global}).
By carrying out the variation on the independent variables
 $w$ and $\overline{w}$, the Euler-Lagrange equations to be solved are:
\[
\frac{d}{d\tau}\frac{\partial{L}}{\partial{ \frac{dw}{d\tau} }}
                                     - \frac{\partial{L}}{\partial{w}} = 0,
\hspace{8mm} \frac{d}{d\tau} \frac{\partial{L}}{\partial{
 \frac{d\overline{w}}{d\tau}
               }} - \frac{\partial{L}}{\partial{\overline{w}}} = 0
\]
which yield the following equation of motion:
\begin{equation}
 m \frac{d^2 w}{d\tau^2} = 0.
\end{equation}
Solving this equation yields solutions of the form
\begin{equation}
 w = \alpha \tau + \beta ,  \label{fps}
\end{equation}
where $\alpha$ and $\beta$ are semi-complex constants.
These solutions are, of course, the straight lines in Minkowski
space (i.e. particles moving with constant velocity).

\section{Local Gauge Invariance}               \label{gravfd}
The dynamics generated by the Lagrangian in (\ref{lag2}) is simple enough ---
free particles in Minkowski space  move with constant velocity.
Moreover, the Lorentz invariance of the theory simply refers to
its invariance under the global phase transformation (\ref{global}).

More interesting dynamics occurs if we make the added assumption that our
theory should remain invariant under a {\em local} phase transformation:
\begin{equation}
\mbox{ \fbox{ $w(\tau) \rightarrow w'(\tau) =
 e^{-j\theta(\tau)}\cdot w(\tau)$} }
 \mbox{local phase invariance} \label{localphase}
\end{equation}
where the phase angle
 $\theta = \theta(\tau)$ is now allowed to vary with $\tau$.
 The invariance demanded
by (\ref{localphase}) is just a generalization of the global
 phase invariance encountered in  (\ref{global}).

Leaping from a global phase invariance to a local one is a very popular
theme in particle physics, and is usually referred to as the
`gauge principle' \cite{Aitchison}.

\medskip
To begin our analysis, we need to  find a suitable
Lagrangian that is preserved by local phase transformations.
The Lagrangian (\ref{lag2})
as it stands, is {\em not} invariant under the local phase transformation
(\ref{localphase}), and so needs to be modified. This can be done
as follows; replace the derivative  $\frac{d}{d\tau}$ by the
 derivative operator
\begin{equation}
 D = \frac{d}{d\tau} - \frac{j}{c^2} g .
\end{equation}
The   gauge field $g$ (which we assume to be real valued)
has the dimensions of acceleration,
and is defined to transform in the following way:
\begin{equation}
g \rightarrow g' = g - c^2 \frac{d\theta}{d\tau}.
\end{equation}
Thus the derivative operator transforms as follows:
\begin{eqnarray*}
D \rightarrow D' & = & \frac{d}{d\tau} + \frac{j}{c^2} g' \\
                 & = & \frac{d}{d\tau} + \frac{j}{c^2} g
                                      -j\frac{d\theta}{d\tau}.
\end{eqnarray*}
These definitions, combined with the local phase transformation
(\ref{localphase}), lead us to the identity
\begin{equation}
D'w' = e^{-j\theta(\tau)} \cdot Dw
\end{equation}
(Exercise!)

If we redefine our Lagrangian to be
\begin{equation}
L = \overline{Dw} \cdot  Dw \label{lagmod}
\end{equation}
then the gauge transformation $w \rightarrow w'$, $g \rightarrow g'$
leaves $L$ above invariant. We should emphasise that the gauge field
$g$, together with its transformation
law,  have to be introduced in order to ensure local phase
invariance.

To see that
 the gauge field $g$ is a suitable candidate for some kind
of {\em force} field,
we invoke relation (\ref{phvel}) to deduce the explicit
form of the transformation law for $g$:
\begin{eqnarray*}
 g \rightarrow g' & = & g - c^2 \frac{d}{d\tau} \tanh^{-1}\frac{v}{c} \\
                  & = & g - \frac{1}{(1-v^2/c^2)^{3/2}} \cdot
                                                 \frac{dv}{dt}.
\end{eqnarray*}
Now  $-(dv/dt)/(1-v^2/c^2)^{3/2}$ corresponds to the {\em relativistic
force} acting on a unit mass (at rest in the fixed frame) as
seen by an observer  moving with velocity $v$ with respect to the
fixed frame. This is precisely the way in which a force field
would
transform in a {\em relativistic} setting.

\medskip
These observations {\em suggest} that the
  gauge group $\{ e^{j\theta} |
\theta \in {\bf R} \}$ might give rise to physical fields.
Our analysis thus far does not offer any clues about the
field equations governing $g$,  but
we will soon remedy this after we introduce the calculus of
several semi-complex variables.

\part{Several Semi-Complex Variables}            \label{severalvariables}

\chapter{Semi-Complex Inner Product Spaces}

\section{Motivation}
Chapter \ref{physics} highlighted some interesting connections that can
arise between semi-complex analysis and special relativity in
one space dimension. However, we live in a space which has manifestly
three spatial dimensions, so we might ask whether our analysis can be
extended in a natural way  to encompass this higher dimensional
case.

Mathematically speaking,
the most natural higher dimensional objects that can be built
from semi-complex numbers  are perhaps {\D}-modules\footnote{
i.e. modules over the ring of semi-complex numbers }, and semi-complex
manifolds. The latter will allow us to discuss the notion of
`curvature' and will be dealt with in Chapter \ref{manifolds}.

There is of course the question of the appropriateness of this theory in
modelling physical reality. The reader is invited to form his or her own
opinion on this one, but it is perhaps safest to view this subject as a kind
of {\em mathematical} curiosity, which admits now and again the possibility
of physical application. In any case, speculations on the relevance of
the theory to mathematical physics will be given free reign in Part
\ref{semiworld}.

The vector spaces ${\bf R}^{n}$ and ${\bf C}^{n}$ should be familiar to the
reader;
we now wish to explore the module space  ${\bf D}^{n}$.
This space admits a very natural inner product that is {\em non-euclidean}
in character. This is the content of the next section.

\section{The Standard Inner Product on ${\bf D}^n$}
The rather elegant properties exhibited by the semi-complex numbers suggest
that they might be useful in constructing higher dimensional objects.
One such object is the module space ${\bf D}^{n}$. Elements of
${\bf D}^{n}$ are the $n \times 1$ matrices\footnote{ or $n$-tuples}
\begin{equation}
w =  \left( \begin{array}{c}
         w^1 \\
         \vdots     \\
         w^n
     \end{array}
 \right) , \hspace{.8cm} w^i \in {\bf D}.
\end{equation}
For each element $w \in {\bf D}^n$, we define $w^{\ast}$ to be the row matrix
\begin{equation}
w^{\ast} = \left( \begin{array}{ccc}
               \cn{w}^1 & \cdots & \cn{w}^n
              \end{array}
        \right)
\end{equation}
i.e. $w^{\ast}$ is just the conjugate transpose of $w$.

\medskip
\noindent
Now consider the map
$< , > : {\bf D}^{n} \times {\bf D}^{n}  \rightarrow  {\bf D}$
given by the rule
\begin{equation}
<w , \omega > =  w^{\ast} \cdot \omega
\label{innerproduct}
\end{equation}
In components, definition (\ref{innerproduct}) takes the form
\begin{eqnarray}
<  \left( \begin{array}{c}
         w^1 \\
         \vdots     \\
         w^n
     \end{array}
   \right),
\left( \begin{array}{c}
         \omega^1 \\
         \vdots     \\
         \omega^n
     \end{array}
 \right) > & = & \left( \begin{array}{ccc}
         \cn{w}^1 &
         \cdots    &
        \cn{w}^n
     \end{array}
 \right) \cdot \left( \begin{array}{c}
         \omega^1 \\
         \vdots     \\
         \omega^n
     \end{array}
 \right)
 \\
& =  & \cn{w}^1 \omega^1 + \cdots + \cn{w}^n \omega^n .
\label{innerexplicit}
\end{eqnarray}
In particular, we have
\begin{equation}
<w,w> = |w^1|^2 + \cdots + |w^n|^2.
\end{equation}
If we write $w^i = t^i + jx^i$, the last expression becomes
\begin{equation}
<w,w> = (t^1)^2 +\cdots + (t^n)^2 - (x^1)^2 - \cdots - (x^n)^2 .
                       \label{metricw}
\end{equation}
So the most `natural' inner product on ${\bf D}^{n}$ induces
a {\em non-euclidean} metric on ${\bf R}^{2n}$. This inner product
will be called the {\em standard inner product} on ${\bf D}^n$.

We  remark that the standard  inner product on ${\bf D}^{n}$ is
{\em non-degenerate}. That is, if $<w,\omega> = 0$ for all $\omega$,
then $w$ must vanish. We will be restricting our investigations
to such non-degenerate cases.

\medskip
\noindent

At this stage, it is interesting to note that if we introduce
a new variable $t$ by writing
\beq
 t^2 = (t^1)^2 + \cdots + (t^n)^2 ,
\eeq
then expression (\ref{metricw}) takes the form
\beq
 <w,w> = t^2 - (x^1)^2 - \cdots - (x^n)^2,
\eeq
which, on its own, induces a {\em Lorentzian} metric
on the real space of $(n+1)$-tuples \[ \{ (t, x^1,\cdots , x^n)\} . \]

\section{Unitary Symmetry on $\Dn$}
Consider the case where ${\bf D}^n$ is equipped with the
standard inner product
defined by (\ref{innerproduct}).
Let $U$ be an $n \times n$ matrix with semi-complex entries, and let $U^{\ast}$
 denote, as usual, the conjugate transpose of $U$. A simple
calculation shows that
\begin{equation}
<Uw, U\omega> = <w,\omega>
\end{equation}
for all $w,\omega \in {\bf D}^n$ if and only if
\begin{equation}
U^{\ast}U = 1.
\end{equation}
Consequently, the set of all (semi-complex) linear transformations
$U: \Dn \rightarrow \Dn$ which
preserve the inner product forms a (non-compact) Lie group. We will
denote this group by U($n,{\bf D}$), which is to be distinguished
from the corresponding complex unitary group U($n, {\bf C}$).

\section{General Inner Product Spaces}  \label{gips}
We are now in a position to list the main properties of the standard
inner product: For all $w_1, w_2, w_3 \in {\bf D}^n$, $\lambda \in {\bf D}$,
we have
\begin{enumerate}
\item $<w_1,w_2> \in {\bf D}$,
\item $<w_1,w_2> = \overline{<w_2,w_1 >}$,
\item $<w_1,\lambda w_2> = \lambda <w_1,w_2>$,
\item $<w_1 + w_2 , w_3 > = <w_1, w_3> + <w_2, w_3>$, and
\item $<,>$ is non-degenerate.
\end{enumerate}
If ${\bf M}$ is any given  module  over {\D}
which admits  a map  $<,>:{\bf M} \times {\bf M} \rightarrow
{\bf D}$ satisfying the properties listed above, it will be called
a {\em semi-complex inner product space}. Obviously, ${\bf D}^n$,
together with the standard inner product, is an example of semi-complex inner
product space. A non-trivial example is given below.

\begin{example} {\em Suppose $H$ is an $n \times n$
{\em hermitian}\footnote{ i.e. $H^{\ast} = H$, where $H^{\ast}$ denotes the
conjugate transpose of $H$.}
 matrix over the semi-complex numbers. Define a map
$ < \cdot | H | \cdot >
 :  {\bf D}^{n} \times {\bf D}^{n}  \rightarrow  {\bf D} $
by writing
\beq
        < w | H | \omega > = w^{\ast} H \omega
 \hspace{7mm}  \forall w,\omega \in \Dn .
\eeq
Then $<\cdot | H | \cdot >$ is a semi-complex inner product whenever
the hermitian matrix $H$ is non-singular. We leave the details to the reader. }
\end{example}

\chapter{The Semi-Complex Unitary Groups}       \label{unitarygroups}
\section{The Hyperbola Group U($1,{\bf D}$)}
If the modulus squared of a semi-complex number  $w$ is unity, $w$
must have the form $e^{j\theta}$ or $-e^{j\theta}$ for some
real parameter $\theta$. Consequently,
\begin{equation}
\mbox{U($1,{\bf D}$)} = \{ \hspace{2mm} \pm e^{j\theta} \hspace{2mm}
 | \hspace{2mm} \theta \in {\bf R} \}.
\end{equation}
Geometrically, U($1,{\bf D}$) is the {\em hyperbola} $t^2-x^2 = 1$. Thus it is
a  non-compact Lie group with two simply connected components,
which are the branches of the hyperbola.
The branch $\mbox{U}_{+}(1, {\bf D}) =
 \{ \hspace{2mm} e^{j\theta} \hspace{2mm} | \hspace{2mm}
\theta \in {\bf R} \}$ is the component of U($1,{\bf D}$) containing
the identity , and in the light
of expression (\ref{lore}), is identified as the restricted group
of Lorentz transformations in $1+1$ space-time. From these definitions,
we deduce that
\begin{equation}
\mbox{U($1,{\bf D}$)}/ \mbox{U}_{+}(1, {\bf D}) \cong {\bf Z}_{2}. \label{cyc1}
\end{equation}
Generalisations of the above identity occur for the higher dimensional
groups, which we begin to investigate next.

\section{The Groups U($2,{\bf D}$) and SU($2,{\bf D}$)}
If $U$ is a  $2 \times 2$ matrix over ${\bf D}$ with the property
\beq
          U^{\ast} U = 1,
\eeq
then
\beq
 \det( U^{\ast} U ) = \cn{ \det U} \cdot \det U = | \det U |^2 = 1.
\eeq
{}From the preceding discussion, we conclude that for such a matrix $U$,
we have, for some real $\theta$, either
\beq
   \det U = +e^{j\theta} \hspace{4mm} \mbox{or} \hspace{4mm}
                                      \det U =  -e^{j\theta}.
\eeq
The set of all elements in U($2,{\bf D}$) for which the first relation
above holds forms a proper subgroup which we denote by
${\mbox{U}}_{+}(2,{\bf D})$. In fact, any element
$U \in {\mbox{U}}_{+}(2,{\bf D})$ can be written in the form
\begin{equation}
U = e^{jH}    \label{Unitary}
\end{equation}
where $H$ is a $2 \times 2$ hermitian matrix over ${\bf D}$
(i.e. $H^{\ast} = H$).

Since ${\mbox{U}}_{+}(2,{\bf D})$ is the connected component of U($2,{\bf D}$)
containing the identity, it is a normal subgroup, and elements
$\sigma$ in the quotient
\beq
\mbox{U($2,{\bf D}$)}/ \mbox{U}_{+}(2, {\bf D})
\eeq
have the property
\beq
       \sigma^2 = 1.
\eeq
The {\em special unitary group} SU($2,{\bf D}$) is defined to be the set
of all elements $U\in$ U($2,{\bf D}$) with the property
\beq
                   \det U = +1.
\eeq
Consider the following traceless, hermitian matrices
\begin{equation}
 \begin{array}{ccc}
\tau_1 = \left( \begin{array}{cc}
                   0 & 1 \\
                   1 & 0
                \end{array}
         \right), &
\tau_2 = \left( \begin{array}{cc}
                    0 & -j \\
                    j & 0
                \end{array}
         \right), &
\tau_3 = \left( \begin{array}{cc}
                   1 & 0 \\
                   0 & -1
                \end{array}
         \right).
  \end{array}
\end{equation}
Then any $2\times 2$ hermitian matrix $H$  may be written in the form
\begin{eqnarray}
 H & = & \theta . 1 + \sum_{k=1}^{3} \phi_k \tau_k \\
   & = & \theta . 1 + \vec{{\bf \phi}}\cdot {\bf \tau},
\end{eqnarray}
where the parameters $\theta, \phi_k$ are real, and $1$ is the $2\times 2$
identity matrix. Utilising the identity
\begin{equation}
\det{ e^{A}} = e^{\mbox{tr}A}, \label{detA}
\end{equation}
we deduce that any element  of the special unitary group SU($2, {\bf D}$)
 may be written in the form
\begin{equation}
U = \exp{(j
\vec{\phi}\cdot \tau /2)}, \hspace{8mm} \phi_k \in {\bf R}
\end{equation}
and that any member of ${\mbox{U}}_{+}(2,{\bf D})$  is a product of an element
in SU($2, {\bf D}$) and a phase $e^{j\theta}$.

The {\em anti}-hermitian matrices\footnote{a matrix $A$ is anti-hermitian if
$A^{\ast} = -A$} $E_{k} = \frac{1}{2}j\tau_k$
form a basis for the {\em real} Lie algebra $su(2,{\bf D})$, and satisfy the
 following commutation
relations:
\begin{eqnarray}
   [E_1,E_2] & = & E_3
\label{com1} \\
  \mbox{[} E_2, E_3] & = & E_1
\label{com2}\\
  \mbox{[} E_3, E_1] & = & -E_2
\label{com3}
\end{eqnarray}
The structure of this Lie algebra becomes clear as soon as we introduce the
new variables $S_{+}, S_{-}, S_3 $ defined
in terms of the $E_k$:
\begin{eqnarray}
 S_{+} & = & j(E_1-jE_2) \\
 S_{-} & = & j(E_1+jE_2) \\
 S_3  & = & jE_3
\end{eqnarray}
{}From  the commutation relations
(\ref{com1})--(\ref{com3}), we deduce the following:
\begin{eqnarray}
 [ S_3 , S_{\pm} ] & = & \pm S_{\pm} \\
 \mbox{[}S_{+} , S_{-} ] & = & 2S_3
\end{eqnarray}
These relations are precisely those given for the real Lie algebra
$sl(2,{\bf R})$ (see, for example, \cite{Sattinger}),
and play a fundamental role in the quantum mechanics of
spin and angular momentum \cite{Schiff}. The
different representations (over ${\bf R}$) for this algebra
yield the quantised values for spin, characterised by the Casimir
operator $S^2 = \frac{1}{2}(S_{+}S_{-} + S_{-}S_{+}) + S_{3}^{2}$.  It can be
shown directly from the commutation relations that the Casimir operator
 commutes with each of the basis elements $S_{+}, S_{-}$ and $S_3$.

We have so far demonstrated that the real Lie algebras $sl(2,{\bf R})$
and $su(2,{\bf D})$ are (semi-complex) isomorphic, but $sl(2,{\bf R})$
is  isomorphic\footnote{strictly speaking, {\em complex } isomorphic;
see \cite{Sattinger}} with $su(2, {\bf C})$
as well. The interesting observation here is that the  quantised values
of spin (or the eigenvalues of the Casimir operator) predicted by
an SU($2,{\bf C})$ theory of spin are identical to an SU($2,{\bf D})$
theory (where SU($2,{\bf D})$ acts on ${\bf D}^2$).

Perhaps this is not too surprising considering the similarities
of the conjugation operation for complex and semi-complex numbers.
It should come as no surprise, then, that the Lie algebras for
SU($3,{\bf C}$) and SU($3,{\bf D}$) are intimately related.

We devote the next section to a (brief) study of U($3,{\bf D}$) and
SU($3,{\bf D}$).

\section{The Groups U($3,{\bf D}$) and SU($3,{\bf D}$)}
If $U$ is a  $3\times 3$ matrix over ${\bf D}$ with the property
\beq
            U^{\ast} U = 1,
\eeq
then we conclude, as before, that for some real $\theta$,
\beq
  \det U = +e^{j\theta} \hspace{4mm} \mbox{or} \hspace{4mm}
                                      \det U =  -e^{j\theta}.
\eeq
The subgroup ${\mbox{U}}_{+}(3,{\bf D})$ is defined to
be the set of all elements in
U$(3,{\bf D})$ with the first property above, and represents the connected
component of U$(3,{\bf D})$ containing the identity. Equivalently, we may write
\begin{equation}
\mbox{${\mbox{U}}_{+}$($3,{\bf D}$)} = \{ \hspace{1mm} e^{jH} \hspace{1mm}
| \hspace{1mm} \mbox{$H$ is $3\times 3$ hermitian} \}.
\end{equation}
The special unitary group SU($3,{\bf D})$ consists of those elements
in U$(3,{\bf D})$ with determinant equal to one.

An element $\sigma$ in the quotient group
$\mbox{U($3,{\bf D}$)}/ \mbox{U}_{+}(3, {\bf D})$ satisfies $\sigma^2 = 1$.

\medskip

In order to determine the Lie algebra of the special unitary
group SU($3,{\bf D}$)
 we use the fact that any $3 \times 3$ hermitian matrix $H$ may be written
in the form
\begin{eqnarray}
H & = & \theta . 1 + \sum_{i=1}^{8} \alpha_{i} \mu_{i} \\
  & = & \theta . 1 + \vec{\alpha} \cdot \mu,
\end{eqnarray}
where the parameters $\theta, \alpha_{i}$ are real, $1$ is the
$3\times 3$ identity matrix, and $\mu_{i}$
are the traceless hermitian matrices listed below:
\small{
\[
\begin{array}{ccc}
\mu_{1} =  \left( \begin{array}{ccc}
                    0 & 1 & 0 \\
                    1 & 0 & 0 \\
                    0 & 0 & 0
                  \end{array}
           \right) &

\mu_{2} =  \left( \begin{array}{ccc}
                    0 & -j & 0 \\
                    j & 0 & 0 \\
                    0 & 0 & 0
                  \end{array}
           \right) &

\mu_{3} =  \left( \begin{array}{ccc}
                    1 & 0 & 0 \\
                    0 & -1 & 0 \\
                    0 & 0 & 0
                  \end{array}
           \right) \\

\mu_{4} =  \left( \begin{array}{ccc}
                    0 & 0 & 1 \\
                    0 & 0 & 0 \\
                    1 & 0 & 0
                  \end{array}
           \right) &
\mu_{5} =  \left( \begin{array}{ccc}
                    0 & 0 & -j \\
                    0 & 0 & 0 \\
                    j & 0 & 0
                  \end{array}
           \right) &
\mu_{6} =  \left( \begin{array}{ccc}
                    0 & 0 & 0 \\
                    0 & 0 & 1 \\
                    0 & 1 & 0
                  \end{array}
           \right) \\

\mu_{7} =  \left( \begin{array}{ccc}
                    0 & 0 & 0 \\
                    0 & 0 & -j \\
                    0 & j & 0
                  \end{array}
           \right) &
\mu_{8} =  \left( \begin{array}{ccc}
                    0 & 0 & 0 \\
                    0 & -1 & 0 \\
                    0 & 0 & 1
                  \end{array}
           \right) &

\end{array}
\] }
\normalsize
With the help of identity (\ref{detA}), one can see that any element $U$
in SU($3,{\bf D}$) may be written in the form
\begin{equation}
 U = \exp{(j\vec{\alpha} \cdot \mu /2 )}
\end{equation}
where the eight parameters $\alpha_{i}$ are real. The anti-hermitian
matrices $F_k = \frac{1}{2}j\mu_k$ are readily seen to form a basis
for the real Lie algebra $su(3,{\bf D})$ , and
satisfy commutation relations
\begin{equation}
 [F_p, F_q] = f_{pqr}F_r. \label{comm3}
\end{equation}
Explicit values for the ({\em real}) structure constants $f_{pqr}$ can easily
be obtained, but in order to illuminate the structure of this
Lie algebra, we introduce new variables $I_3, Y, I_{\pm}, U_{\pm}, V_{\pm}$,
defined in terms of the $F_k$:
\begin{eqnarray*}
I_3 & = & jF_3 \\
Y & = & 2j(\frac{1}{3}F_3 - \frac{2}{3}F_8) \\
 I_{\pm} & = & j(F_1 \mp jF_2) \\
 U_{\pm} & = & j(F_6 \mp jF_7) \\
 V_{\pm} & = & j(F_4 \mp jF_5).
\end{eqnarray*}
{}From these definitions, and the commutation relations (\ref{comm3}),
 we deduce the following:
\[
\begin{array}{cc}
\mbox{[}I_3, I_{\pm} ] = \pm I_{\pm}  & \mbox{[}Y, I_3 ] = 0 \\
\mbox{[}I_3, U_{\pm} ] = \mp \frac{1}{2} U_{\pm} &
\mbox{[}Y, U_{\pm} ] = \pm U_{\pm} \\
\mbox{[}I_3, V_{\pm} ] = \pm \frac{1}{2} V_{\pm} &
\mbox{[}Y, V_{\pm} ] = \pm V_{\pm}
\end{array}
\]
Incidentally, these commutation relations played a significant role
in a formal quark theory of matter
 (\cite{Close},
\cite{Sattinger})
dealing with the up, down and strange quarks\footnote{now known to be
related to quark `flavours'}. In this  `flavour' theory of quarks,
 $I_3$ and $Y$ are isospin
and hypercharge operators, while  $U_{\pm}$, $V_{\pm}$, and $I_{\pm}$,
 are raising
and lowering operators.

Our observations show that the  the
predicted values for isospin and hypercharge in the
old flavour  theory of quarks (where the symmetry group
SU($3,{\bf C}$) acts on ${\bf C}^3$) can equally be obtained
by   an  SU($3,{\bf D}$) theory, where elements in SU($3,{\bf D}$)
act on ${\bf D}^3$.

\medskip
In quantum mechanics, the complex unitary groups arise from considering the
`rotational' symmetries of a vector wave function {\bf $\Psi$}, which
has complex components. At this stage, it is tempting to ask if a vector
valued function with semi-complex valued components has any
potential application to physics. In other words, can an element in
${\bf D}^3$, say, represent some physical state of a system, or of space-time?

\chapter{Semi-Complex Manifolds and Curvature} \label{manifolds}
 Fortunately, the subject of semi-complex manifolds can be
handled in precisely the same way as for complex manifolds.
Our  initial discussion follows
very closely the one given in  \cite{Nakahara}.
 Other possible sources that may be
helpful to the reader include \cite{Field} and \cite{Felice}.

\section{Semi-Complex Manifolds}
To begin, we  define a holomorphic map on $\Dm$. A semi-complex valued function
$f: \Dm \rightarrow {\bf D}$ is {\em holomorphic} if
\beq
              \frac{\db f}{\db \cn{w}^{\mu}} = 0
\eeq
for $\mu = 1, \dots m$ ($w^{\mu} = t^{\mu} + jx^{\mu}$). A map
$(f^1,\dots ,f^n) : \Dm \rightarrow \Dn$ is called holomorphic if each function
$f^{\lambda}$ ($1 \leq \lambda \leq n$) is holomorphic.

To define  a semi-complex manifold, we invoke the
usual definition given for a complex manifold, except
we assume the transition functions are {\em semi-complex} holomorphic
instead of complex holomorphic.

In fact, for most of this chapter, the reader may safely arrive
at the correct formal definitions
(which we will often omit for brevity) by referring to the
corresponding complex case, and replacing the phrase
`complex holomorphic' with the alternative phrase
`semi-complex holomorphic'.

\begin{definition} {\em
 $M$ is a semi-complex manifold if
 \begin{enumerate}
  \item $M$ is a topological space;
  \item $M$ is provided with a family of pairs
        $\{ (U_i , \phi_i) \}$ such that  $\{ U_i \}$
        is an open cover of $M$, and each $\phi_i$ is a
        homeomorphism from $U_i$ to an open subset of $\Dm$;
  \item Given $U_i$ and $U_j$ with non-empty intersection,
        the transition function $\psi_{ij} = \phi_j \circ \phi^{-1}_{i}$
        from $\phi_i (U_i \cap U_j)$ to $\phi_j (U_i \cap U_j)$
        is semi-complex holomorphic.
 \end{enumerate}
            }
\end{definition}
In words, a semi-complex manifold is a geometrical object which
{\em locally} looks like $\Dm$. The number $m$ is called the
semi-complex dimension of $M$ and we often write
$\mbox{dim}_{\bf D}M = m$.

\section{Calculus on Semi-Complex Manifolds}
The reader is reminded that our discussion is intended to be
brief, since the terminology we use is identical in form
to the one used in the treatment of complex manifolds.

\subsection{The Holomorphic Tangent Module}
Let $M$ be a semi-complex manifold with $\mbox{dim}_{\bf D}M = m$.
Take a point $p$ in a chart $(U,\phi)$ of $M$. The $m$
semi-complex linear operators
\[
\left\{ \frac{\db}{\db w^1}, \dots , \frac{\db}{\db w^m} \right\} \]
 at the point $p$ generate a module space over \D.
We  call it the  holomorphic tangent module at $p$, and denote it by
$T_p M^{+}$. By convention, any element $W \in T_p M^{+}$ will be
called a holomorphic vector.

The anti-holomorphic tangent module $T_p M^{-}$ is the module
generated by the anti-linear operators
\[
\left\{ \frac{\db}{\db \cn{w}^1}, \dots , \frac{\db}{\db \cn{w}^m} \right\}, \]
and any element in $T_p M^{-}$ will be  called an anti-holomorphic vector.

\subsection{Hermitian Manifolds}
The form $dw^{\mu}$ will be viewed as a semi-complex linear map
\[ dw^{\mu} : T_p M^{+} \rightarrow {\bf D} \]
with the property
\[ dw^{\mu} ( \frac{\db}{\db w^{\nu}} ) = \delta^{\mu}_{\nu} . \]
We may now introduce the anti-linear map
\[  d\cn{w}^{\mu} : T_p M^{+} \rightarrow {\bf D} \]
by writing
\[ d\cn{w}^{\mu}(W) = \cn{dw^{\mu}(W)} \]
for any given holomorphic vector $W$.

\medskip

The tangent module $T_p M^{+}$ may be viewed as a semi-complex inner product
space if there exists a map
\[ g: T_p M^{+} \times T_p M^{+} \rightarrow {\bf D} \]
satisfying, for any $W,X,Y \in T_p M^{+}$, $\lambda \in {\bf D}$, the following
properties:
\begin{enumerate}
 \item $g(X,Y) = \cn{g(Y,X)}$;
 \item $g(X,\lambda Y) = \lambda g(X,Y)$;
 \item $g(W,X+Y) = g(W,X)+g(W,Y)$;
 \item $g$ is non-degenerate (see below).
\end{enumerate}
The first three properties imply that $g$ has the form
\beq
 g = g_{\cn{\mu} \nu} d\cn{w}^{\mu} \otimes dw^{\nu},
\eeq
where the semi-complex components $g_{\cn{\mu} \nu}$ transform tensorially.
The last property of non-degeneracy means the matrix $\{g_{\cn{\mu} \nu} \}$
is invertible.

If we now write $<X,Y> = g(X,Y)$, then the map $<,>$ is obviously
a semi-complex inner product on the tangent module $T_p M^{+}$
(refer to Section \ref{gips}).

\medskip

If at each point on a manifold there exists such a map, the
manifold is said to be {\em Hermitian}, and the map $g$
is called a {\em Hermitian metric}. Justifying this terminology
is simple enough; the property $g(X,Y) = \cn{g(Y,X)}$ implies
\beq
              \cn{g_{\cn{\mu} \nu}} = g_{\cn{\nu} \mu} ,
\eeq
and so the matrix  $\{g_{\cn{\mu} \nu} \}$ must be Hermitian.

\subsection{Covariant Derivatives}
Let $M$ be a semi-complex manifold, and suppose $X$ is a holomorphic
vector field on $M$ (i.e. $X(p) \in T_p M^{+}$ for each $p\in M$).
If we assume that any holomorphic vector $V \in T_p M^{+}$ parallely
transported (by some connection map $\Gamma$) to another point
$q$ is again a holomorphic vector $\tilde{V} \in T_q M^{+}$, then the
covariant derivative of $X$ takes the form
\beq
 \nabla_k X^i = \frac{\db X^i}{\db w^k} + \Gamma_{jk}^{i} X^j,
\eeq
where the field of numbers $\Gamma_{jk}^{i}$ are known as
connection coefficients.

\subsection{Metric Compatibility} \label{mcompatibility}
If we now endow the manifold $M$ with a Hermitian metric $g$,
there turns out to be a natural way in which to uniquely
specify the value of these coefficients. We simply impose the
condition that the connection preserves the inner product induced by
the metric $g$. In particular, if two holomorphic vectors
$V$ and $W$ at the point $p$ are `parallel transported' into
the vectors $\tilde{V}$ and $\tilde{W}$ at $q$ (respectively),
then we always have the identity
\[ <V,W>_p = <\tilde{V}, \tilde{W}>_q , \]
where we have used the notation $<X,Y> = g(X,Y)$.

One can show that this
{\em metric compatibility condition} takes the equivalent form
\beq
     \nabla_k g_{\cn{i} j} = 0 .
\eeq
Evaluating the covariant derivative of the metric $g$ in the last
expression yields the condition
\beq
 \frac{\db g_{\cn{i} j}}{\db w^k} - g_{\cn{i} n} \Gamma^n_{jk} = 0 .
\eeq
The connection coefficients are thus uniquely determined, since we can
express them in terms of the components of the metric:
\beq
 \Gamma^i_{jk} = g^{i \cn{n}} \frac{\db g_{\cn{n} j}}{\db w^k} ,\label{cuniq}
\eeq
where $\{g^{n \cn{m}} \}$ is the inverse matrix of $\{ g_{\cn{i} j} \}$.
Such a connection is called a {\em Hermitian connection}.

\section{The Curvature}
Our aim in this section is to  list the main results
concerning the curvature of a semi-complex manifold.
A thorough treatment can be found (at least in form)
in  \cite{Nakahara} and \cite{Felice}.
\subsection{Torsion}               \label{torsionsec}
The torsion tensor $T$ is defined by writing
\beq
                    T^i_{jk} = \Gamma^i_{jk} - \Gamma^i_{kj}.
\eeq
For a Hermitian connection, it takes the form
\beq
         T^i_{jk} = g^{i \cn{n}} \left\{
       \frac{\db g_{\cn{n} j}}{\db w^k} -
       \frac{\db g_{\cn{n} k}}{\db w^j} \right\} .
\eeq
So if the torsion vanishes, the metric components satisfy the following
relations:
\beq
      \frac{\db g_{\cn{n} j}}{\db w^k} =
       \frac{\db g_{\cn{n} k}}{\db w^j}.       \label{kalmet}
\eeq
A Hermitian metric $g$ satisfying these equations is called
a {\em K\"{a}hler metric}.

\subsection{Geodesics}
A {\em geodesic} is a curve ${\gamma}$ whose tangent vector
is everywhere non-zero and proportional to a
parallely propagated vector. Intuitively, it is the `straightest'
possible curve.

In a particular coordinate system, the geodesic $\gamma$ has a parametrization
$w(\tau) = (w^1(\tau), \dots ,w^n(\tau))$ which can be
shown to satisfy (see \cite{Felice}) the geodesic equation
\beq
 \frac{d^2 w^i}{d\tau^2} + \Gamma^i_{jk} \frac{dw^j}{d\tau}
                                             \frac{dw^k}{d\tau} = 0,
          \label{geq}
\eeq
for some affine parameter $\tau$.

As in Riemannian geometry, the geodesic equation may be derived
from an action principle whenever the metric is {\em torsion free}
(or K\"{a}hler). In fact, if we choose as our Lagrangian the expression
\beq
            L = g_{\cn{\mu} \nu} \frac{d\cn{w}^{\mu}}{d\tau} \cdot
                                 \frac{dw^{\nu}}{d\tau},
\eeq
then the Euler-Lagrange equations can be shown to give the
geodesic equation (\ref{geq})
whenever $g$ is K\"{a}hler.

\subsection{The Riemann Tensor}
We will now state the main results concerning the form
of the curvature tensor (or {\em Riemann Tensor} $R$) for
a semi-complex manifold. Again, for a complete explanation
of the concepts involved, the reader should refer to
\cite{Nakahara}, \cite{Felice} or \cite{Ward}.

\medskip

Suppose $M$ is a semi-complex manifold with a connection
$\Gamma$. Then the components of the Riemann tensor satisfy
the following relations:
\begin{eqnarray}
 R^j_{irm} & = & \db_r \Gamma^j_{im} - \db_m \Gamma^j_{ir} +
      \Gamma^n_{im} \Gamma^j_{nr} - \Gamma^n_{ir} \Gamma^j_{nm} ;
                                     \label{Ra} \\
 R^j_{i \cn{r} m}  & = & \db_{\cn{r}} \Gamma^j_{im} ;
                                     \label{Rb} \\
 R^j_{ir \cn{m}} & = & - \db_{\cn{m}} \Gamma^j_{ir}. \label{Rc}
\end{eqnarray}
In the above expressions, we used the symbol $\db_r$ to denote the operator
$\frac{\db}{\db w^r}$, and likewise, $\db_{\cn r}$ is shorthand for
writing $\frac{\db}{\db \cn{w}^r}$.

{}From expressions (\ref{Rb}) and (\ref{Rc}), we deduce the identity
\beq
             R^j_{im \cn{r}} = -R^j_{i \cn{r} m}
\eeq
For a Hermitian connection on $M$ (see Section \ref{mcompatibility}), a nice
cancellation occurs; namely, we  deduce the following:
\beq
                  R^j_{irm} = 0.
\eeq
({\em Hint}: substitute(\ref{cuniq}) into (\ref{Ra})).
So the only non-vanishing components of the Riemann tensor
are the relatively simple expressions
 $R^j_{i \cn{r} m}$ and  $R^j_{ir \cn{m}}$.

\subsection{The  Ricci Form}            \label{ricciforms}
New quantities of interest may be obtained by contracting indices of
the Riemann tensor. To begin, let us write
\beq
 {\cal{R}}_{r \cn{m}} = R^i_{ir \cn{m}} ( = -\db_{\cn{m}}\Gamma^i_{ir} ).
\eeq
For a Hermitian connection, a convenient simplification occurs\footnote{
namely, we use the result $ \db_r \ln \det g =
g^{i \cn{n}} \db_r g_{\cn{n} i} $. }, and we may write
\beq
 {\cal{R}}_{r \cn{m}} = -\db_r \db_{\cn{m}} \ln \det g ,
\eeq
where $g =\{ g_{\cn{\mu} \nu} \}$. See \cite{Nakahara} for more details.

Since the determinant of a Hermitian matrix is real, $\det g$
is a real quantity. Consequently,
\beq
\cn{{\cal{R}}_{\mu \cn{\nu}}} = \cn{-\db_{\mu} \db_{\cn{\nu}} \ln \det g}
           = -\db_{\nu} \db_{\cn{\mu}} \ln \det g = {\cal{R}}_{\nu \cn{\mu}},
\eeq
and so the matrix of elements $\{{\cal{R}}_{\mu \cn{\nu}} \}$ forms a Hermitian
matrix.

To define the {\em Ricci form}, we write
\beq
  {\cal{R}} = j {\cal{R}}_{\mu \cn{\nu}} dw^{\mu} \wedge d\cn{w}^{\nu},
\eeq
which is a {\em real form}, since $\cn{\cal{R}} = \cal{R}$.

\subsection{The Ricci Tensor}  \label{Riccit}
The Ricci tensor $Ric$  is also obtained by a contraction of
indices:
\beq
 Ric_{r \cn{m}} = R^i_{ri \cn{m}} = -\db_{\cn{m}} \Gamma^i_{ri} .
\eeq
For a Hermitian connection with vanishing torsion (i.e. for
a K\"{a}hler metric),
\beq
 Ric_{r \cn{m}} = -\db_{\cn{m}}\Gamma^i_{ri} = -\db_{\cn{m}} \Gamma^i_{ir}
                 = {\cal{R}}_{r \cn{m}},
\eeq
and so the components of the Ricci form agree with $Ric_{r \cn{m}}$. So
for a K\"{a}hler metric, we have
\beq
 \mbox{ \fbox{$ Ric_{\mu \cn{\nu}} = -\db_{\mu} \db_{\cn{\nu}} \ln \det g $}.}
\eeq
 If
$Ric = {\cal{R}} = 0$, the K\"{a}hler metric is said to be
{\em Ricci-flat}.  In this case, finding solutions to the equation
\beq
 Ric_{\mu \cn{\nu}} = 0   \label{ricciflat}
\eeq
amounts to solving
\beq
\mbox{ \fbox{ $\db_{\mu} \db_{\cn{\nu}} \ln \det g = 0$}}. \label{626}
\eeq
Further insights are gained by solving this equation on a local
co-ordinate chart. Firstly, by virtue of the vanishing torsion condition
expressed by equation (\ref{kalmet}), we may write {\em locally} the expression
\beq
              g_{\cn{\mu} \nu} = \db_{\cn{\mu}} \db_{\nu} \phi, \label{kmmm}
\eeq
where $\phi$ is some scalar valued function. Since $g$ is hermitian, $\phi$
must be {\em real} valued. Now equation (\ref{626}) implies that
\beq
     \ln \det g = F(w^1,\dots,w^n) + \cn{F}({\cn{w}}^1,\dots,{\cn{w}}^n),
       \label{627}
\eeq
where $F$ is an arbitrary semi-complex holomorphic function in the variables
$w^1,\dots,w^n$. Here, $n$ is the (semi-complex) dimension of the manifold.

Performing a local co-ordinate transformation, we can arrange for $F$ to
have the constant value $j$, which means equation (\ref{627}) becomes
\beq
                \det g = 1.
\eeq
Making use of the expression
$g_{\cn{\mu} \nu} = \db_{\cn{\mu}} \db_{\nu} \phi$, we have
\beq
\mbox{ \fbox{ $ \det\{\db_{\cn{\mu}} \db_{\nu} \phi \} = 1$}}.
\eeq
So, locally, solving for a Ricci flat metric $g$ involves choosing
a real scalar $\phi$ which satisfies this last condition.

Whether there exist topological obstructions preventing the existence
of a globally defined metric of this kind is an open question.

In the complex case, the necessary and sufficient condition
for the existence of a unique Ricci flat metric is known; namely,  the
first Chern class, $c_1$, must vanish. See \cite{Nakahara}.

\part{Living in a Semi-Complex World}  \label{semiworld}
\chapter{The Arrow of Time}
\section{A Semi-Complex View of Space-Time}
Our analysis of the semi-complex number system in Part \ref{onevariable}
yielded various results that seemed to be closely related with the theory
of special relativity in one space dimension. For example, if we replace
the space-time point $(t,x)$ by the {\em single} semi-complex variable
$w=t+jx$ (we have chosen units so that $c=1$), we can effect a Lorentz
transformation by multiplying $w$ by a phase factor $e^{-j\theta}$.

At first glance, it is not at all clear how one can apply
 our semi-complex
formalism to the case of four dimensional space-time.
 In particular, is there a natural representation
of the real $4$-tuple,
\beq
 (t,x^1,x^2,x^3), \label{fourspace}
\eeq
($t$ is time; the $x^i$'s are spatial coordinates)
in terms of semi-complex quantities?

Given any space-time point $(t,x^1,x^2,x^3)$, we can define an associated
element
$w\in {\bf D}^3$ by writing
\beq
  w = \left( \begin{array}{r}
                  t+jx^1 \\
                    jx^2 \\ \label{vector4}
                    jx^3
             \end{array}
      \right) .
\eeq
With the standard inner product $<,>$ on ${\bf D}^3$, we have
\begin{eqnarray}
    <w,w> & = & w^{\ast} \cdot w \\
          & = & |t+jx^1|^2 + |jx^2|^2 + |jx^3|^2 \\
          & = & t^2 - (x^1)^2 - (x^2)^2 - (x^3)^2 . \label{lorgg}
\end{eqnarray}
Preserving this last quantity under coordinate transformations is then related
to the problem of finding transformations on ${\bf D}^3$ which preserve
the standard inner product.

One such transformation which preserves the form of (\ref{vector4}) is the
matrix
\beq
 U = \left( \begin{array}{ccc}
            e^{-j\theta} & & \\
              & 1 & \\
             & & 1
             \end{array}
      \right),
\hspace{4mm} \theta \in {\bf R}.
\eeq
Note that $U^{\ast} U = 1$, and $\det U = e^{-j\theta}$, which implies
that $U \in {\mbox{U}}_{+}(3,{\bf D})$. This corresponds to a velocity boost
in the direction of the $x^1$ axis.

The reader may have noticed that the representation of a general
space-time point (\ref{fourspace}) by the column vector
(\ref{vector4}) is slightly lopsided, since the time variable $t$ appears in
the first row, and not in the second or third. Of course, we would still end
up with the result (\ref{lorgg}) if the time variable appeared in the second
or third row. In fact, a completely symmetrical representation can be achieved
by `distributing' the time variable amongst the three rows.

More precisely, suppose
 $t^1,t^2$ and $t^3$ are any real numbers satisfying
\beq
 \mbox{ \fbox{$(t)^2 = (t^1)^2 + (t^2)^2 + (t^3)^2$},} \label{loree}
\eeq
and let us define
\beq
    w = \left( \begin{array}{r}
                  t^1+jx^1 \\
                  t^2+jx^2 \\  \label{sixspace}
                  t^3+jx^3
             \end{array}
      \right).
\eeq
Then
\begin{eqnarray}
 <w,w> & = & |t^1 +jx^1|^2 + |t^2+jx^2|^2 + |t^3+jx^3|^2 \\
       & = & (t^1)^2 + (t^2)^2 +(t^3)^2 - (x^1)^2-(x^2)^2 -(x^3)^2 \\
       & = & t^2 - (x^1)^2-(x^2)^2 -(x^3)^2.
\end{eqnarray}
Again, preserving this last quantity is related to finding transformations
on ${\bf D}^3$ which preserve the standard inner product $<,>$.

Now any element of ${\mbox{U}}_{+}(3,{\bf D})$ acting on (\ref{sixspace})
will preserve the inner product $<,>$, and so it induces (by virtue
of the relation (\ref{loree})) a transformation on Minkowski space
$\{ (t, x^1,x^2,x^3) \}$ which preserves the Lorentz metric on this space.

Incidentally,  this induced transformation on
Minkowski space is
{\em non-linear}.

\medskip

We now consider infinitesimal displacements of space and time.
In special relativity, the space-time metric is flat and has the form
\beq
     ds^2 = dt^2 - (dx^1)^2 - (dx^2)^2 - (dx^3)^2 .
\eeq
If we write
\beq
        dw = \left( \begin{array}{r}
                  dt+jdx^1 \\
                     jdx^2 \\ \label{dvector4}
                     jdx^3
             \end{array}
      \right) ,
\eeq
then we can recover the space-time metric as follows:
\begin{eqnarray}
      <dw,dw> & = & dw^{\ast} \cdot dw \\
              & = & dt^2 - (dx^1)^2 - (dx^2)^2 - (dx^3)^2.
\end{eqnarray}
A more symmetrical formulation can be achieved by introducing the
quantities
$dt^1,dt^2,dt^3$ which are assumed to satisfy the identity
\beq
\mbox{ \fbox{$(dt)^2 = (dt^1)^2+(dt^2)^2+(dt^3)^2$}.} \label{dtrule}
\eeq
Then, setting
\beq
        dw = \left( \begin{array}{r}
                  dt^1+jdx^1 \\
                  dt^2+jdx^2 \\ \label{dsymm}
                  dt^3+jdx^3
             \end{array}
      \right),
\eeq
we can recover the (flat) space-time metric in the usual way:
\begin{eqnarray*}
     <dw,dw> & = & (dt^1)^2+(dt^2)^2+(dt^3)^2
                       - (dx^1)^2 - (dx^2)^2 - (dx^3)^2 \\
             & = &  dt^2 - (dx^1)^2 - (dx^2)^2 - (dx^3)^2.
\end{eqnarray*}

\section{Time's Arrow}
Notice that we can write (\ref{sixspace}) in the form
\beq
              w = \vec{t} +j\vec{x},
\eeq
where
\beq
      \vec{t} = \left( \begin{array}{c}
                           t^1 \\
                           t^2 \\                         \label{timevector}
                           t^3
                       \end{array}
                 \right) \in {\bf R}^3
\eeq
is called the {\em time vector}, and
\beq
      \vec{x} = \left( \begin{array}{c}
                           x^1 \\
                           x^2 \\
                           x^3
                       \end{array}
                 \right) \in {\bf R}^3
\eeq
is the usual {\em position vector}.

The time $t$ (or {\em physical time}) is simply the {\em length}
of the time vector:
\beq
            t = | \vec{t} | .
\eeq
Similarly, we may write (\ref{dsymm}) in the form
\beq
              dw = \vec{dt} + j\vec{dx},
\eeq
where $\vec{dt}$ is the infinitesimal time displacement vector, and
$\vec{dx}$ is the infinitesimal position displacement vector.

The physical time displacement $dt$ is given by the length
of the infintesimal $\vec{dt}$:
\beq
          dt = | \vec{dt} | .
\eeq

\medskip

So far we have attempted to symmetrise the representation of
a space-time point (or infinitesimal space-time displacement) in terms of
our semi-complex formalism by introducing a vector quantity
 whose length specifies the usual time variable $t$ (or displacement $dt$).
However, this prescription does not tell us what particular element of the form
(\ref{sixspace}) corresponds to the space-time point $(t,x^1,x^2,x^3)$,
since the identity (\ref{loree}) does not uniquely define the variables
$t^1, t^2,t^3$.

On the other hand, we can assign to each element in ${\bf D}^3$
a unique\footnote{strictly speaking, there is a sign ambiguity here.}
 space-time point by invoking this same identity. Formally
speaking, we have
 an onto map from ${\bf D}^3$ to Minkowski space:
\beq
    \left( \begin{array}{r}
                  t^1+jx^1 \\
                  t^2+jx^2 \\
                  t^3+jx^3
             \end{array}
      \right)
        \longrightarrow
      (t,x^1,x^2,x^3),
\eeq
where $t$ is obtained
 by invoking identity (\ref{loree}).
In this sense, ${\bf D}^3$ may be viewed as an `augmentation' of
Minkowski space; or, equivalently, Minkowski space may be viewed
as a `derived' space. In any case, we choose to work in the space
${\bf D}^3$ since the mathematics appears to be formally
more elegant. This is in large part due to the extra symmetry afforded
by introducing three time variables which we can match with the
three well known space variables, allowing us to work in
{\em three} semi-complex dimensions, rather than four real dimensions.

For now, we should look upon the introduction of a `time vector'
as a pure mathematical convenience. Speculations concerning
the physical significance of our formalism will be discussed
later.

\chapter{General Relativity on Semi-Complex Manifolds}
\pagestyle{myheadings}
\markright{GR ON SEMI-COMPLEX MANIFOLDS}
In this chapter we briefly examine Einstein's General
Theory of Relativity in the context of curved semi-complex
manifolds. In particular, we investigate equation
(\ref{ricciflat}) for Ricci-flat (semi-complex ) space-times.

As a preliminary, we introduce the {\em wave equation}
for the semi-complex space-time ${\bf D}^3$.
\section{The Wave Equation on ${\bf D}^3$}       \label{secwave}
Let $\Phi: {\bf D}^3 \rightarrow {\bf D}$ be a semi-complex valued
scalar field on ${\bf D}^3$, and set
\beq
        \eta = \left( \begin{array}{c}
                       \frac{\db \Phi}{\db w^1} \\
                       \frac{\db \Phi}{\db w^2} \\
                       \frac{\db \Phi}{\db w^3}
                      \end{array}
               \right) .
\eeq
If we define the  Lagrangian density  $\cal{L}$ by writing
\begin{eqnarray}
 \cal{L} & = & \eta^{\ast} \cdot \eta \\
     & = & \frac{\db \cn{\Phi}}{\db {\cn{w}}^1} \cdot \frac{\db \Phi}{\db w^1}
         +\frac{\db \cn{\Phi}}{\db {\cn{w}}^2} \cdot \frac{\db \Phi}{\db w^2}
         +\frac{\db \cn{\Phi}}{\db {\cn{w}}^3} \cdot \frac{\db \Phi}{\db w^3},
\end{eqnarray}
then the equation of motion\footnote{i.e. the Euler-Lagrange equations}
for $\Phi$ is
\beq
 \left( \frac{\db^2}{\db {\cn{w}}^1 \db w^1} +
            \frac{\db^2}{\db {\cn{w}}^2 \db w^2}                 \label{wave3}
        +\frac{\db^2}{\db {\cn{w}}^3 \db w^3} \right) \Phi = 0 ,
\eeq
where
\beq
   \frac{\db^2}{\db {\cn{w}}^{\mu} \db w^{\mu}} =
   \frac{\db}{\db {\cn{w}}^{\mu}} \cdot
     \frac{\db}{\db  w^{\mu}} =                              \label{op3}
     \frac{1}{4} \left\{ \frac{\db^2}{\db (t^{\mu})^2} -
        \frac{\db^2}{\db (x^{\mu})^2} \right\} , \hspace{4mm} \mu = 1,2,3.
\eeq
As a matter of convenience, we choose to define
\beq
 \Box_{\mu} =   \frac{\db^2}{\db {\cn{w}}^{\mu} \db w^{\mu}} .
\eeq
Adopting this notation enables us to rewrite (\ref{wave3}) as
\beq
 ( \Box_{1} + \Box_{2} + \Box_{3} ) \Phi = 0 .
\eeq
Of course, we can generalise our definition of the wave equation
for the space ${\bf D}^n$. Namely, the $n$-dimensional semi-complex
wave equation is defined by
\beq
     \left( \sum_{\mu = 1}^{n} \Box_{\mu} \right) \Phi = 0 .
\eeq
We will be primarily concerned with the three dimensional case.

\medskip

By virtue of relations (\ref{op3}), the wave equation (\ref{wave3})
can be put into the more explicit form
\beq
 \frac{1}{4} \left(
     \frac{\db^2}{\db (t^1)^2}+ \frac{\db^2}{\db (t^2)^2}
      + \frac{\db^2}{\db (t^3)^2} -  \frac{\db^2}{\db (x^1)^2}
       -  \frac{\db^2}{\db (x^2)^2} -  \frac{\db^2}{\db (x^3)^2} \right)
          \Phi = 0 .
\eeq
Recalling that the (physical) time variable $t$ is determined by the length
of the time vector (\ref{timevector}), we may write
\begin{eqnarray}
    t^1 & = & t\cos \theta \sin \phi \\
    t^2 & = & t \sin \theta \sin \phi \\
    t^3 & = & t \cos \phi
\end{eqnarray}
where we have chosen the usual {\em spherical coordinates}
$\theta$ and $\phi$ to specify the direction of the time vector.
The variable $t$ is then the radial coordinate determined by the distance
of the point $(t^1,t^2,t^3) \in {\bf R}^3$ from the origin.

In terms of these spherical coordinates, the operator
\beq
  \frac{\db^2}{\db (t^1)^2}+ \frac{\db^2}{\db (t^2)^2}
      + \frac{\db^2}{\db (t^3)^2}
\eeq
becomes
\beq
 \frac{1}{t^2} \frac{\db}{\db t} \left( t^2 \frac{\db }{\db t} \right)
  + \frac{1}{t^2 \sin^2 \phi } \frac{\db^2 }{\db \theta^2} +
  \frac{1}{t^2 \sin \phi} \frac{\db}{\db \phi}
   \left(\sin \phi \frac{\db }{\db \phi} \right) .          \label{wavesph}
\eeq
 To say the function $\Phi$ is dependent only on the physical
time $t$ and the spatial coordinates $x^1,x^2,x^3$ implies the
condition
\beq
      \frac{\db \Phi}{\db \theta} = \frac{\db \Phi}{\db \phi} = 0 .
\eeq
This last condition is tantamount to saying that $\Phi$ is spherically
symmetric with respect to the time co-ordinates $t^{\mu}$, and enables
us to view $\Phi$ as a well defined function on Minkowski space.

For this spherically symmetric case, the wave equation on ${\bf D}^3$
takes the form
\beq
   \frac{1}{t^2} \frac{\db}{\db t} \left( t^2 \frac{\db \Phi}{\db t} \right)
       -    \frac{\db^2 \Phi}{\db (x^1)^2}
       -  \frac{\db^2 \Phi}{\db (x^2)^2} -
       \frac{\db^2 \Phi}{\db (x^3)^2}      \label{wav}
           = 0 .
\eeq
If we now make the substitution
\beq
    \Phi = \frac{\Psi}{t}, \hspace{6mm} t\neq 0,   \label{subgr}
\eeq
equation (\ref{wav}) becomes (for $t \neq 0$)
\beq
   \frac{\db^2 \Psi}{\db t^2}  -    \frac{\db^2 \Psi}{\db (x^1)^2}
       -  \frac{\db^2 \Psi}{\db (x^2)^2} -
       \frac{\db^2 \Psi}{\db (x^3)^2}      \label{wavemink}
           = 0 ,
\eeq
which is the standard wave equation on Minkowski space.

\section{Ricci-Flat Space-Times}
Ricci-flat space-times in Einstein's General Theory of relativity
are obtained by solving the equation
\beq
                   R_{ij} = 0,
\eeq
where $R_{ij}$ is the usual Ricci tensor defined for a real
four dimensional manifold.

For semi-complex manifolds, the analogous equation to be solved is
\beq
           Ric_{\mu \cn{\nu}} = 0 , \label{richy}
\eeq
where the Ricci tensor $Ric$ for semi-complex manifolds
was introduced in section \ref{ricciforms}. We also showed that in the case
of a Hermitian connection with vanishing torsion, equation (\ref{richy})
assumes the relatively simple form
\beq
            \db_{\mu} \db_{\cn{\nu}} \ln G = 0 , \label{richyex}
\eeq
where $G$ denotes the determinant of the Hermitian matrix
 $\{ g_{\cn{\mu} \nu } \}$.

In one semi-complex dimension, this equation is simply
\beq
     \frac{\db}{\db {\cn{w}}^1} \cdot \frac{\db}{\db w^1} \ln g_{\cn{1}1} =
 \frac{1}{4} \left\{ \frac{\db^2}{\db (t^1)^2} -
        \frac{\db^2}{\db (x^1)^2} \right\} \ln g_{\cn{1}1} = 0,
\eeq
and so $\ln  g_{\cn{1}1}$ satisfies the one dimensional wave equation.
Therefore, the general solution for equation (\ref{richyex}) is simply
\beq
        g_{\cn{1}1} = e^{2\sigma},
\eeq
where $\sigma$ is a solution of the one dimensional wave equation.

Incidentally, this  is equivalent to the analytical solution one would obtain
from Einstein's equation (\ref{richy})  for a real
{\em two} dimensional manifold. Of course, all real two dimensional manifolds
are conformally flat, so our result does not say anything profound in
a geometrical sense, but it does indicate that our formalism
yields a correct {\em analytical} description of Einsteinian gravity for
the low dimensional case, and so there is a suggestion that
for  higher dimensions, we may also obtain a correct description of
gravity.

\medskip

For semi-complex manifolds of higher dimension, however,  equation
(\ref{richyex}) is {\em non-linear}. In fact, we saw in Section \ref{Riccit}
that, locally, a Ricci flat K\"{a}hler metric  is given by
\beq
     g_{\cn{\mu} \nu} = \db_{\cn{\mu}} \db_{\nu} \phi,
\eeq
where $\phi$ is a real valued scalar function satisfying the equation
\beq
                     \det \{\db_{\cn{\mu}} \db_{\nu} \phi \} = 1.
\eeq
It is thus a challenge to show that in {\em three} semi-complex dimensions,
we can find a solution $\phi$ that gives rise to a geometry corresponding
to the Schwarzschild solution of general relativity (after suitably
projecting the dynamics from the semi-complex space to a real four-manifold).
Presently, the author knows of no such solution.

In any case, we can gain considerable insight by investigating
approximate solutions for the case of `weak curvature'.
We take up this topic next.

\subsection{The Weak Curvature Approximation}   \label{approxcurve}
Our aim in this section is essentially to linearise equation
(\ref{richyex}). We also choose to work in {\em three} semi-complex
dimensions, although our method of approach generalises easily
to other dimensions.

For the case of weak curvature, we expect the matrix of metric coefficients
$g = \{ g_{\cn{\mu} \nu} \}$ to deviate only slightly from the identity
matrix. So if we write
\beq
      g_{\cn{\mu} \nu} =  \delta_{\cn{\mu} \nu} + h_{\cn{\mu} \nu},
                   \label{gravexp}
\eeq
then the Hermitian matrix $h = \{ h_{\cn{\mu} \nu} \}$ measures just
this deviation, and the entries $h_{\cn{\mu} \nu}$ may be assumed to be
 small order
quantities (i.e. $h_{\cn{\mu} \nu} \ll 1$).

Retaining only quantities which are first order in $h_{\cn{\mu} \nu}$,
we have
\beq
      G \equiv \det( g_{\cn{\mu} \nu}) \approx 1 +  h_{\cn{1} 1} + h_{\cn{2} 2}
                                                  + h_{\cn{3} 3},
\eeq
and so
\begin{eqnarray}
        \ln G & \approx & \ln( 1 +  h_{\cn{1} 1} + h_{\cn{2} 2}
                                                  + h_{\cn{3} 3}) \\
              & \approx &   h_{\cn{1} 1} + h_{\cn{2} 2}
                                                  + h_{\cn{3} 3} \\
              & = & \mbox{trace}(h).
\end{eqnarray}
Substituting this last result into equation (\ref{richyex})
yields the following equations:
\begin{eqnarray}
         \Box_{\mu} \mbox{trace($h$)} & = & 0 \hspace{4mm} \mu = 1,2,3.
                                   \label{trace3} \\
 ( \Box_1 + \Box_2 + \Box_3) h_{\cn{\mu} \nu} & = &
                                                 0 \hspace{4mm} \mu \neq \nu.
                \label{box3}
\end{eqnarray}
Equation (\ref{box3}) is obtained by making repeated use of the identities
\beq
  \frac{\db h_{\cn{\mu} \nu}}{\db w^{\lambda}} =
     \frac{\db h_{\cn{\mu} \lambda}}{\db w^{\nu}} ,      \label{twoid}
       \hspace{6mm} \cn{ h_{\cn{\mu} \nu}} = h_{\cn{\nu} \mu},
\eeq
which follow from the fact that $g$ is a K\"{a}hler metric
(refer to Section \ref{torsionsec}).

Equation (\ref{box3}) is immediately identified as the three dimensional
semi-complex wave equation, which we saw was closely linked to the
standard wave equation on four dimensional space-time. In particular,
for the case where these fields are spherically symmetric with respect
to time (and so well defined on Minkowski space),
we conclude  from (\ref{subgr}) and (\ref{wavemink}) that the weak
curvature solutions represent waves which propagate at constant speed,
and are {\em modulated by the amplitude {\em $1/t$}}.

\subsection{Newtonian Gravity}
We now wish to consider the case where the metric components
are {\em time independent}. In particular, we analyse the geodesics for
a stationary, Ricci-flat\footnote{we allow for the possibility that
the Ricci tensor may not be defined at some spatial point}
 (semi-complex) space-time in the weak curvature approximation.

As in Section \ref{approxcurve}, the matrix $h = \{ h_{\cn{\mu} \nu} \}$
will measure the (small) deviation away from flat space-time.  The
condition of time independence for the metric components $g_{\cn{\mu} \nu}$
implies (by virtue of relation (\ref{gravexp}))
\beq
            \frac{\db  h_{\cn{\mu} \nu}}{\db t^{\lambda}} = 0 ,
                \hspace{6mm} \lambda = 1,2,3.
\eeq
Hence, we may write
\beq
   \frac{\db  h_{\cn{\mu} \nu}}{\db w^{\lambda}} =
    \frac{1}{2} j   \frac{\db  h_{\cn{\mu} \nu}}{\db x^{\lambda}} .
\eeq
The first identity in (\ref{twoid}) now takes the form
\beq
    \frac{\db  h_{\cn{\mu} \nu}}{\db x^{\lambda}} =
      \frac{\db  h_{\cn{\mu} \lambda}}{\db x^{\nu}}. \label{kahlersp}
\eeq
Combining this last result with the Hermitian condition for $h$
(second identity in (\ref{twoid})) enables one to prove in three dimensions
that the matrix  $h = \{ h_{\cn{\mu} \nu} \}$ is {\em real symmetric}.
The proof is slightly involved, but  the interested
reader is strongly encouraged to go through the details!
Since the components of the matrix $h$ are now real, we will write
$h = \{ h_{\mu \nu} \}$. So the real-symmetric property of $h$
translates into the identity
\beq
          h_{\mu \nu} = h_{\nu \mu} . \label{rsymm}
\eeq

\medskip
Since the components  $h_{\mu \nu}$ are time independent, equations
(\ref{trace3}) and (\ref{box3}) reduce to
\begin{eqnarray}
    \frac{\db^2}{\db (x^{\mu})^2} \mbox{ trace($h$)} & = & 0 \hspace{4mm}
                                                 \mu = 1,2,3;
                 \label{tind} \\
    \nabla^2 h_{\mu \nu} & = & 0 \hspace{4mm} \mu \neq \nu, \label{tind2}
\end{eqnarray}
where $\nabla^2$ is the usual (three dimensional) Laplacian.
Also, to first order in small quantities,
the connection coefficients $\Gamma^{\mu}_{\lambda \nu}$ defined by
equation (\ref{cuniq}) are given by
\beq
        \Gamma^{\mu}_{\lambda \nu} \approx \frac{1}{2} j
          \frac{\db h_{\mu \lambda}}{\db x^{\nu}} =
   \frac{1}{2} j \frac{\db h_{\lambda \nu}}{\db x^{\mu}} ,
\eeq
where in the last step we used the relations (\ref{kahlersp}) and
(\ref{rsymm}).

For this weak curvature approximation, the geodesic equation (\ref{geq})
becomes
\beq
 \frac{d^2 w^{\mu}}{d \tau^2} +
   \frac{1}{2} j \frac{\db h_{\lambda \nu}}{\db x^{\mu}}
  \frac{dw^{\lambda}}{d\tau} \frac{dw^{\nu}}{d\tau} = 0 . \label{geqa}
\eeq
The invariant parameter $\tau$ is the `proper time', and is connected
with the physical time $t$ through the relation
\beq
             d\tau^2  =  g_{\cn{\mu} \nu}
             \frac{d{\cn{w}}^{\mu}}{dt} \frac{dw^{\nu}}{dt} dt^2 .
\eeq
We will now assume that the velocity $(dx^1/dt, dx^2/dt, dx^3/dt)$
which can be calculated from the parametrization $w(\tau) =
(w^1(\tau), w^2(\tau),w^3(\tau))$ using identity (\ref{dtrule}), has
magnitude much less than the speed of light (which in our present
units is one). In General Relativity, this is known as taking
the non-relativistic or `Newtonian' limit. With this assumption, we
 may replace $\tau$ by $t$
in the geodesic equation (\ref{geqa}) to obtain
\beq
 \frac{d^2}{dt^2}(t^{\mu} +jx^{\mu}) +
   \frac{1}{2} j \frac{\db h_{\lambda \nu}}{\db x^{\mu}}
  \frac{dt^{\lambda}}{dt} \frac{dt^{\nu}}{dt} = 0 ,
    \hspace{5mm} \mu = 1,2,3.
\eeq
Equating real and imaginary parts, we get
\begin{eqnarray}
     \frac{d^2 t^{\mu}}{dt^2} & = & 0 ;    \label{timec} \\
     \frac{d^2 x^{\mu}}{dt^2} & = &   -\frac{1}{2}
  \frac{\db h_{\lambda \nu}}{\db x^{\mu}}
  \frac{dt^{\lambda}}{dt} \frac{dt^{\nu}}{dt}  \hspace{7mm} \mu = 1,2,3.
               \label{accel}
\end{eqnarray}
The reader may notice that equation (\ref{accel}) gives the actual acceleration
of a point particle defined by the geodesic.

Let us define the variables $k^1,k^2,k^3$ by writing
\beq
                k^i = \frac{dt^i}{dt}, \hspace{5mm} i=1,2,3.
\eeq
Notice from equation (\ref{timec}) that the $k^i$'s are {\em constant},
and moreover, satisfy the identity $(k^1)^2 + (k^2)^2 +(k^3)^2 =1$.

\medskip

If we now introduce the potential function
\beq
          \phi = \frac{1}{2} h_{\mu \nu} k^{\mu} k^{\nu},
                      \label{pot}
\eeq
we may rewrite (\ref{accel}) as the familiar expression
\beq
            \vec{a} = -\nabla \phi,   \label{veca}
\eeq
where $\vec{a}$ denotes the acceleration vector.

{}From the definition of the potential $\phi$, and the time independent
equations (\ref{tind}) and (\ref{tind2}), we deduce the identity
\beq
 \nabla^2 \phi = \frac{1}{2} \left[
     (k^1)^2 \nabla^2 h_{11} +
      (k^2)^2 \nabla^2 h_{22} +  \label{mass}
        (k^3)^2 \nabla^2 h_{33} \right].
\eeq
{}From the Newtonian viewpoint, the right hand side of this equation
represents a mass source term, which in this case is induced
by the diagonal elements of the matrix $h$, and the time component
derivatives $k^i$. Our theory therefore provides a mechanism  for
obtaining sources of mass from a massless field. Notice, however,
that for the symmetrical case $k^1=k^2=k^3$, this source term vanishes
(by an application of equation (\ref{tind})). In this case, the potential
satisfies
\beq
                    \nabla^2 \phi = 0 .
\eeq
Combining this result with relation (\ref{veca}),
we conclude that for the radially symmetric case
(allowing for a singularity at the origin),
 the acceleration is governed by an inverse
square law. Of course, this is the same conclusion obtained
by (a rather involved) analysis of Einstein's equations.

\pagestyle{headings}
\chapter{${\mbox{U}}_{+}(1,{\bf D})$ Gauge Theory}
\section{Introductory Remarks}
Two very fundamental forces in nature that have been known
for a long time are the electromagnetic force (giving rise
to the attraction or repulsion between two charged particles)
and the force of gravity. The first can be described by a very elegant
mathematical theory known as gauge field theory. In particular,
the underlying gauge group for electromagnetic interactions
is the circle group
\beq
      \mbox{U}(1,{\bf C}) = \{ e^{i\theta}| \theta \in {\bf R} \}.
\eeq
The electro-weak force, and the proposed interactions between quarks,
are also modelled using gauge theory (see \cite{shaw},\cite{love}).
For an excellent introduction into this subject, the reader is referred
to \cite{Aitchison}.

In all of these cases, the underlying gauge group is relatively simple in form.
Electro-weak theory is based on the {\em complex} gauge group SU$(2) \times$
U$(1)$, while interactions between quarks seem to satisfy the SU$(3)$
gauge. It is a curious fact that all of these gauge groups
(including the one for the electromagnetic interaction) are (or are
composed of) the {\em complex unitary groups}.

Perhaps this is not too surprising if one recalls that the language of
quantum mechanics is deeply rooted in the Hilbert space formalism,
in which unitary  operators enjoy the property of preserving the inner
product.

\medskip

We saw in Section \ref{gravfd} that the semi-complex  group
\beq
        {\mbox{U}}_{+}(1,{\bf D}) = \{ e^{j\theta}| \theta \in {\bf R} \}
\eeq
may give rise to physical fields consistent with special relativity.
We now take up this topic by first discussing ${\mbox{U}}_{+}(1,{\bf D})$
gauge invariance for a massless scalar field.

\section{The Massless Scalar Field}   \label{massless}
The Lagrangian density
\beq
       {\cal{L}} =
          \frac{\db \cn{\Phi}}{\db {\cn{w}}^1} \cdot \frac{\db \Phi}{\db w^1}
         +\frac{\db \cn{\Phi}}{\db {\cn{w}}^2} \cdot \frac{\db \Phi}{\db w^2}
         +\frac{\db \cn{\Phi}}{\db {\cn{w}}^3} \cdot \frac{\db \Phi}{\db w^3}
\eeq
first discussed in Section \ref{secwave} is invariant under the {\em global}
phase transformation
\beq
    \Phi \rightarrow \Phi' = e^{j\theta} \cdot \Phi, \hspace{6mm} \theta =
                                                         \mbox{constant},
\eeq
and gives rise to the wave equation on ${\bf D}^3$. We now invoke the
`gauge principle' by imposing the condition of
{\em local} phase invariance---or invariance under the transformation
\beq
     \Phi \rightarrow \Phi' = e^{j\theta(w)} \cdot \Phi, \label{opps}
\eeq
where $\theta= \theta(w)$ is now allowed to vary for different space-time
points $w\in {\bf D}^3$.

 We first need to modify our Lagrangian density
$\cal{L}$ in order to satisfy the invariance demanded by (\ref{opps}).
We start by introducing the derivative operators
\beq
      D_{\mu} = \frac{\db}{\db w^{\mu}} + j \Omega_{\mu}
                                       \hspace{6mm} \mu = 1,2,3,
\eeq
where the gauge field  $\Omega = \{ \Omega_{\mu} \}$ obeys the
folowing transformation law:
\beq
        \Omega_{\mu} \rightarrow \Omega'_{\mu} = \Omega_{\mu} -
                              \frac{\db \theta}{\db w^{\mu}}. \label{tom}
\eeq
The operator now transforms as
\beq
  D_{\mu} \rightarrow D'_{\mu} = \frac{\db}{\db w^{\mu}} + j
                                          \Omega'_{\mu}.
\eeq
With this terminology, it is a straightforward exercise to show that
\beq
         D'_{\mu} \Phi' = e^{j\theta(w)} \cdot D_{\mu} \Phi.
\eeq
We may now redefine our Lagrangian density to be
\beq
 {\cal{L}} = \cn{D_1 \Phi} \cdot D_1 \Phi + \cn{D_2 \Phi} \cdot D_2 \Phi
                  + \cn{D_3 \Phi} \cdot D_3 \Phi ,  \label{llinv}
\eeq
which is invariant under the gauge transformations $\Phi \rightarrow \Phi'$
and $\Omega_{\mu} \rightarrow \Omega'_{\mu}$.

\medskip

The
equation of motion determined by this Lagrangian density
is:
\beq
    ( \Box_{\ast} + | \Omega |^2 ) \Phi   =  -V \Phi,    \label{higgs}
\eeq
where $ \Box_{\ast}= \Box_1 + \Box_2 + \Box_3$,  $|\Omega |^2
= |\Omega_1|^2
+|\Omega_2|^2 + |\Omega_3|^2$, and
\beq
    V = j\left( \Omega_{\mu} {\cn{\db}}_{\mu} + {\cn{\Omega}}_{\mu} {\db}_{\mu}
                  +  {\cn{\db}}_{\mu} {\Omega}_{\mu} \right).
\eeq
We have also used the notation ${\cn{\db}}_{\mu} \equiv {\db}_{\cn{\mu}}$,
and summed over repeated indices.

\medskip

So far, we have said nothing about the likely equations of motion governing
the field $\Omega$. We address this topic next.

\section{The ${\mbox{U}}_{+}(1,{\bf D})$ Field Tensor}
The gauge group
${\mbox{U}}_{+}(1,{\bf D}) = \{ e^{j\theta}| \theta \in {\bf R} \}$ and
transformation law (\ref{tom}) for the gauge field $\Omega$ is analogous
to the gauge theory formalism employed in a treatment of the
electromagnetic field. We now strengthen the analogy by defining
the quantity
\beq
 {\cal{F}}_{\mu \cn{\nu}} = j\left( \frac{\db \Omega_{\mu}}{\db {\cn{w}}^{\nu}}
                    -  \frac{\db {\cn{\Omega}}_{\nu}}{\db w^{\mu}} \right) ,
\eeq
which is analogous to the electromagnetic field tensor. We will call
$\cal{F}$ the ${\mbox{U}}_{+}(1,{\bf D})$ {\em field tensor}.

The main reason for defining $\cal{F}$ in this way is that it is invariant
under the gauge transformation $\Omega_{\mu} \rightarrow \Omega'_{\mu}$
defined by equation (\ref{tom}) (easy exercise!). Notice that
\beq
        \cn{ {\cal{F}}_{\mu \cn{\nu}}} =  {\cal{F}}_{\nu \cn{\mu}},
\eeq
which means the matrix $\{ {\cal{F}}_{\mu \cn{\nu}} \}$ is Hermitian, and so
\beq
           j {\cal{F}}_{\mu \cn{\nu}} dw^{\mu} \wedge d{\cn{w}}^{\nu}
\eeq
is a {\em real} form.

\medskip

In order to obtain the Lagrangian density for the gravitational field,
we contract indices by writing
\beq
   {\cal{L}} =  {\cal{F}}_{\mu \cn{\nu}} {\cal{F}}^{\cn{\nu} \mu},
\eeq
where indices are raised using the Kronecker delta $\delta^{\cn{\nu} \mu}$,
and lowered using $\delta_{\mu \cn{\nu}}$.

The Euler-Lagrange equations for this Lagrangian yield the following field
equations:
\beq
  \mbox{\fbox{$ \db_{\mu} {\cal{F}}^{ \hspace{2mm} \mu}_{\nu} = 0$}},
                   \hspace{6mm} \nu = 1,2,3,
\eeq
where $\db_{\mu} \equiv \frac{\db}{\db w^{\mu}}$.
In terms of the fields $\Omega_{\mu}$, this last equation
becomes\footnote{as usual, we sum over repeated indices}
\beq
 \mbox{ \fbox{ $\Box_{\ast} \Omega_{\mu} - \db_{\mu}(\db_{\nu}
                           {\cn{\Omega}}_{\nu}) = 0 $}} \hspace{4mm}
                                         \mu = 1,2,3. \label{grod}
\eeq
Since $\Omega$ is  a semi-complex valued vector field, we may write
\beq
     \Omega_{\mu} = G_{\mu} + jH_{\mu},   \hspace{6mm} \mu = 1,2,3,
\eeq
where $G= \{ G_{\mu} \}$ and $H = \{ H_{\mu} \}$ are real  fields.

For the case where the fields are {\em independent of time},
equations (\ref{grod})
reduce to
\begin{eqnarray}
     \nabla^2 G + \nabla(\nabla \cdot G) & = & 0 \label{trone} \\
      \nabla^2 H - \nabla(\nabla \cdot H) & = & 0 \label{crcrh}
\end{eqnarray}
Invoking a well known identity\footnote{ $\nabla  \times ( \nabla
 \times A ) = \nabla ( \nabla \cdot A ) - \nabla^2 A $ },
equation (\ref{crcrh}) is
equivalent to
\beq
           \nabla {\bf \times} ( \nabla {\bf \times} H ) = 0, \label{maxwell}
\eeq
which is analogous to  Maxwell's field equation for the vector potential
in the static case. In fact, making use of the gauge freedom, we can
choose a gauge in which $\nabla \cdot H$ vanishes, and so
we end up with Laplace's equation for $H$. We {\em cannot}
do the same for the $G$ field, since the gauge transformation for $G$
involves only time coordinates.

Let us rewrite (\ref{trone}) in the form
\beq
        2\nabla (\nabla \cdot G ) -
            \nabla {\bf \times} ( \nabla {\bf \times} G ) = 0.
\eeq
Thus, the only spherically symmetric, radially directed (i.e. curless)
solution to the above equation is
\beq
              G = -\Lambda_0 \vec{r} - \frac{M_0 \vec{r}}{r^3},
\eeq
where $\vec{r} = (x^1,x^2,x^3)$ (with length $r$), and $\Lambda_0,M_0$ are
constants. If we view $G$ as some kind of force field, then the
above solution corresponds to Hooke's Law and Newton's inverse square Law.

It is interesting to note that only these force laws give rise to stable,
closed orbits of particles. See \cite{goldstein} for more details!

\section{The Road Ahead}
Perhaps the most compelling reason for wishing to study
space-time and gravity within a semi-complex
framework is the striking similarity between the
semi-complex unitary groups (arising from a natural
investigation of the symmetries of ${\bf D}^n$) and
the complex unitary groups that underly the modern
description of particle interactions.

In particular, the complex gauge group SU(3) is believed
to model quark interactions, while the semi-complex
gauge group SU(3, ${\bf D}$) preserves the inner product
on the semi-complex space-time ${\bf D}^3$. Are these two
facts related?

If any significant progress is to be made, we should at least
start by investigating the gauge fields arising
from the group SU$(3,{\bf D})$, and applying
some kind of quantisation procedure.  Presumably, we should expect
a new kind of physics, but hopefully, not too unlike the
world inhabited by quarks and gluons!



\end{document}